\documentclass[review]{elsarticle}

\usepackage{hyperref}

\journal{Physics Reports}

\usepackage{amsmath}
\usepackage{amssymb}
\usepackage{graphicx}
\usepackage{color}
\usepackage{xspace}
\usepackage{nicefrac}
\usepackage[dvipsnames]{xcolor}
\newcommand\myshade{85}
\usepackage[T1]{fontenc}
\usepackage{lmodern}
\usepackage{tablefootnote}
\usepackage{ulem}
\usepackage{floatflt}
\usepackage{wrapfig}

\hypersetup{
 citecolor = mycitecolor!\myshade!black,
 linkcolor = mylinkcolor!\myshade!black,
 colorlinks = true,
}

\def\be{\begin{equation}}
\def\ee{\end{equation}}
\def\ba{\begin{eqnarray}}
\def\ea{\end{eqnarray}}

\begin{document}

\begin{frontmatter}

%\shorttitle{Merging stellar-mass binary black holes}    

%\shortauthors{Mandel \& Farmer}  

\title{Merging stellar-mass binary black holes}
%\shorttitle{}

\author[1,2,3]{Ilya Mandel}
%\email{ilya.mandel@monash.edu}
%\cormark
\ead{ilya.mandel@monash.edu}
\affiliation[1]{organization={Monash Centre for Astrophysics, School of Physics and Astronomy, Monash University}, city={Clayton, Victoria}, citysep={}, postcode={3800}, country={Australia}}
\affiliation[2]{organization={OzGrav, ARC Centre of Excellence for Gravitational Wave Discovery}, \country={Australia}}
\affiliation[3]{organization={Institute of Gravitational Wave Astronomy and School of Physics and Astronomy, University of Birmingham}, city={Birmingham}, citysep={}, postcode={B15 2TT}, \country={United Kingdom}}
\author[4]{Alison Farmer}
\affiliation[4]{organization={kW Engineering}, city={Oakland}, state={California}, postcode={94612}, \country={USA}}

\begin{abstract}
The LIGO and Virgo detectors have directly observed gravitational waves from mergers of pairs of stellar-mass black holes, along with a smaller number of mergers involving neutron stars.  These observations raise the hope that compact object mergers could be used as a probe of stellar and binary evolution, and perhaps of stellar dynamics. This colloquium-style article summarises the existing observations, describes theoretical predictions for formation channels of merging stellar-mass black-hole binaries along with their rates and observable properties, and presents some prospects for gravitational-wave astronomy.
\end{abstract}

%\begin{keyword}
%Doh
%\end{keyword}

%\maketitle

\end{frontmatter}

%\linenumbers

\section{Introduction}

Our goal is to present an informal, colloquium-style article on the subject of merging stellar-mass black-hole binaries as gravitational-wave sources. Several reviews were written in the excitement following the first detection \citep{GW150914} of gravitational waves by the Laser Interferometer Gravitational-wave Observatory (LIGO) \citep{GW150914:detectors}, including the astrophysical context companion paper by the the LIGO-Virgo scientific collaboration \citep{GW150914:astro} and the excellent article by \citet{Miller:2016}.  More recent contributions included reviews by \citet{Mapelli:2018,Mapelli:2021} and a compilation of compact binary merger rate measurements and predictions by \citet{MandelBroekgaarden:2021}. However, it is challenging to formally review a field that is both rapidly evolving and still very much in its infancy.  With around 70 detections of gravitational waves from merging black holes \citep{BBH:O1,BBH:O2,Venumadhav:2020,GWTC2,Abbott:2021-GWTC-2-1,GWTC3,Nitz:2021-4OGC,Olsen:2022}, along with two from merging neutron stars \citep{GW170817, GW190425} and two from mergers of neutron stars with black holes \citep{GW200105}, we are only beginning to explore the population of merging compact-object binaries.  The understanding of the formation channels for these sources is evolving with each new data release.

The full range of astrophysical questions that gravitational-wave observations will answer is itself a topic of active study and debate. For example: What can gravitational-wave observations of merging binaries tell us about the stellar and binary evolution that preceded the mergers?  Can the observations constrain the amount of mass loss and expansion experienced by massive stars? Or the stability and consequences of mass transfer, including the infamous common-envelope phase of evolution? Or the amount of mass fallback during supernova explosions, or the kicks that supernovae impart to compact objects? Does the redshift distribution of merging compact objects contain an imprint of the star formation history of the Universe?  Can we use these mergers to probe dynamics in dense stellar environments such as globular clusters? The answers to these important and exciting questions may also in turn influence our understanding of topics as diverse as reionisation and heavy-element nucleosynthesis.

In this article, we focus on the questions that strike us as being most interesting and timely at this stage of the field's development.  The choice of these questions is undoubtedly biased by our own interests. Further, we do not attempt to systematically survey all of the relevant literature.  In other words, this is emphatically {\it not} a review \citep{magritte}. 

We summarise the existing observations of merging binary black holes and closely related systems in \autoref{obs}, describe the plausible formation scenarios for binary black holes in \autoref{form}, discuss the predicted merger rates and merging object properties in \autoref{merge}, and speculate about the prospects for gravitational-wave astronomy in \autoref{prospect}.

\section{Observations}\label{obs}

\subsection{Gravitational-wave observations}

\subsubsection{The confirmed detections to date}
During its first observing run, lasting from September of 2015 through January of 2016, the advanced LIGO detector network observed three signals from the mergers of binary black holes:  GW150914 \citep{GW150914}, GW151226 \citep{GW151226} and GW151012 \citep{GW150914:rates,BBH:O1}.  During the second observing run, lasting with some interruptions from November 2016 through August 2017, three further confident binary black hole detections were announced: GW170104 \citep{GW170104},  GW170608 \citep{GW170608}, and GW170814 \citep{GW170814}, the last of these with the participation of the Virgo gravitational-wave observatory \citep{AdvVirgo}.  The first half of the third observing run, which lasted from April to October of 2019, saw detections of nearly 40 binary black hole mergers \citep{GWTC2,Abbott:2021-GWTC-2-1}, among which GW190412 \citep{GW190412}, GW190814 \citep{GW190814}, GW190521 \citep{GW190521} stood out for their mass and mass ratio measurements (see section \ref{sec:GWevents}).  Approximately 30 more binary black holes (the exact numbers here and above depend on the assumed threshold for confident detections) were observed during the second half of the third observing run, from November 2019 through March 2020 \citep{GWTC3}.  Additional candidates were found by search pipelines external to the LIGO-Virgo-KAGRA collaborations in LIGO-Virgo public data \citep{Nitz:2019,Nitz:2021,Venumadhav:2019,Venumadhav:2020,Nitz:2021-4OGC,Olsen:2022}.  

Meanwhile, the detection of gravitational waves from the binary neutron star merger GW170817 \citep{GW170817} was followed up by an unprecedented campaign of electromagnetic observations, leading to detections of a short gamma ray burst, optical kilonova, and an afterglow spanning from radio through X-ray wavelengths \citep{GW170817:GRB,GW170817:MMA}.  Another confident detection of a binary neutron star merger, GW190425 \citep{GW190425}, as well as two mergers of neutron stars with black holes,  GW200105 and GW200115 \citep{GW200105}, did not yield electromagnetic counterparts.  
We shall focus on binary black holes for the remainder of this article.

\subsubsection{Information encoded in gravitational-wave signals}
The gravitational-wave signature encodes the properties of the merging binary black holes: the component masses and spins (see, e.g., \cite{Schmidt:2020} for a review of gravitational waveform modelling, a topic we will not touch on here). Coupled with information from two or more detectors in a network, this makes it possible to infer the sky location and orientation of the source and the distance to the source \citep{Veitch:2014,GW150914:PE, Ashton:2019}.  

However, some of the source parameters are strongly correlated, leading to near-degeneracies when attempting to extract them from a noisy dataset.   For the heaviest black-hole binaries such as GW150914 and GW190521, the total mass $M \equiv M_1 + M_2$ is measured relatively accurately because the latest stage of the merger waveform, the ringdown, whose frequency is a function of the total mass, falls in the sensitive frequency band of current gravitational-wave detectors.  For lower-mass black-hole binaries, the chirp mass $M_c \equiv M_1^{3/5} M_2^{3/5} M^{-1/5}$ is the better measured parameter, since it determines to lowest order the rate of frequency evolution during the earlier inspiral phase of the waveform.  The mass ratio $q\equiv M_2/M_1 \leq 1$ is often quite poorly constrained, because it enters the gravitational-wave phase evolution during inspiral at a higher order in the ratio of the orbital velocity to the speed of light and is partially degenerate with the black-hole spins \citep[e.g.,][]{PoissonWill:1995,Hannam:2013}.  

\subsubsection{Information extracted from the observed gravitational-wave signals}\label{sec:GWevents}

\begin{figure}
	\centering
	\includegraphics[width=0.99\textwidth]{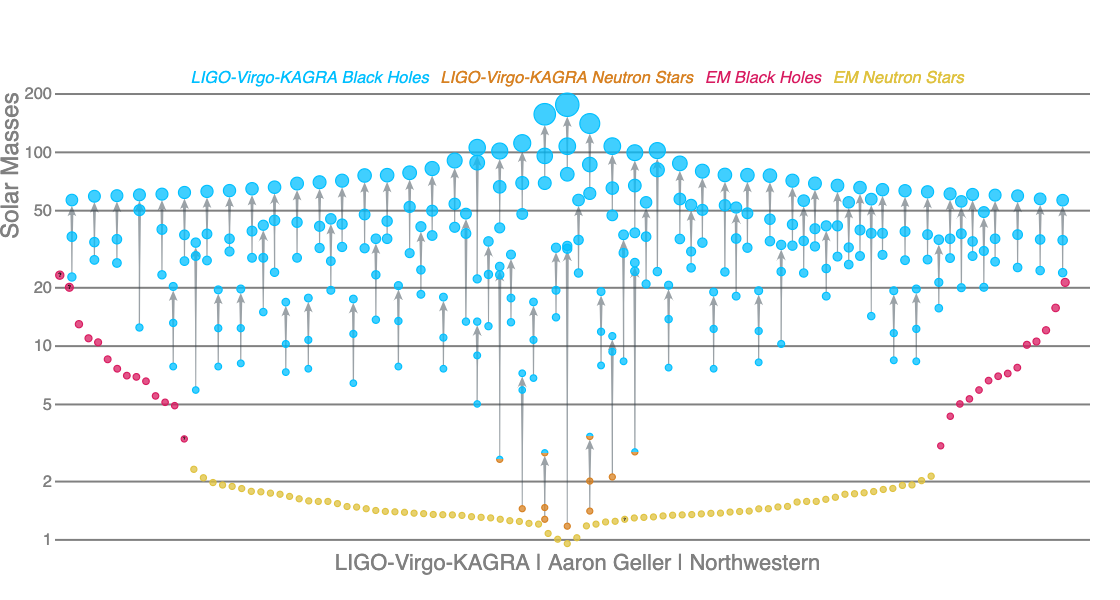}
	\caption{The masses (in solar masses, vertical axis) of black holes observed as gravitational-wave sources and reported in the GWTC-3 catalog (blue), as well as masses of black holes in X-ray binaries (red).  Galactic neutron stars observed as radio pulsars (yellow) and the neutron star candidates in merging compact binaries (orange) appear at the bottom of the figure.  Merger product masses are shown for gravitational-wave detections. Placement on the horizontal axis is arbitrary.  Figure courtesy of Aaron Geller, Northwestern University and LIGO-Virgo-KAGRA collaborations.}\label{fig:BHmasses} 
\end{figure}

The masses of the black-hole binaries observed to date are shown in figure \ref{fig:BHmasses}, along with masses inferred from X-ray binary observations. Specific system properties are discussed below.

\textbf{Masses:} The individual black-hole masses span a range from $\simeq 2.6 M_\odot$ (assuming the lower-mass object in GW190814 is a black hole rather than a neutron star \citep{GW190814}) to $\simeq 80 M_\odot$ for GW190521 \citep{GW190521} (or more, see \cite{FishbachHolz:2020,NitzCapano:2021}) and GW200220\_061928 \citep{GWTC3}.  There is a significant selection bias  toward detecting more massive systems, because sensitivity to gravitational waves from binary mergers depends on the system mass.  The gravitational-wave amplitude, and hence the maximum (horizon) distance at which a source is detectable, scales as $M_c^{5/6}$ for inspiral-dominated signals. The surveyed volume therefore scales as $M_c^{2.5}$.  Once selection effects are accounted for, the mass distribution for the more massive black hole appears to be consistent with a power-law $p(M) \propto M^{-3.4}$, reminiscent of the stellar initial mass function \citep{Salpeter:1955}, with an additional peak at around $34 M_\odot$.  The vast majority of merging binary black holes, $\approx 99\%$, have both companion masses below $45 M_\odot$ \citep{GWTC3:pop}.   

Most observed merging binary black holes are consistent with having equal-mass components, though the poorly measured mass ratios span a very broad range in some cases.  Several observed events, such as GW190412 with a mass ratio of $3:1$ \citep{GW190412}, and GW190814 with a mass ratio of $9:1$ \citep{GW190814}, are definitely inconsistent with equal masses.

\textbf{Black-Hole Spins:} Individual black-hole spins are difficult to measure precisely with gravitational waves.  Most observed mergers are consistent with negligible spins, but some events show evidence of non-zero `effective spin' $\chi_\textrm{eff}$ --- the mass-weighted dimensionless spin along the direction of the orbital angular momentum.  The spin distribution is a topic of active ongoing discussion \citep[e.g.,][]{GWTC2:pop,Roulet:2021,Callister:2021,Galaudage:2021,GWTC3:pop}.  In particular, \citet{Galaudage:2021} conclude that $\sim 80\%$ of merging binary black holes have negligible spins on both components (consistent with zero), while the remaining 20\% have positive and preferentially aligned spins, with no observed events displaying a definite negative $\chi_\textrm{eff}$.  The degree of precession due to spin-orbit misalignment is also unclear; there may be some evidence for it in individual events, such as GW190521 \citep{GW190521} and GW200129 \citep{Hannam:2021}, though other explanations for their signal morphology (e.g., eccentricity, see below) have been proposed.

\textbf{Distances and Sky Locations:} All of the observed binary black hole signals came from median redshifts between $z\sim 0.05$ and $z \sim 0.9$ (distances of around 250 Mpc to 6 Gpc), consistent with the detector sensitivity during the first three observing runs of advanced LIGO \citep{GWTC3}.  With just two operational detectors, LIGO Hanford and LIGO Livingston, for all events prior to GW170814, sources could only be localised to 90\% credible regions spanning hundreds to more than a thousand square degrees on the sky.  The participation of the Virgo instrument in the observation of GW170814 reduced the sky region to only 60 deg$^2$ \citep{GW170814}.   The best-localised binary black hole mergers have 90\% credible sky localisations of only $\approx 20$ deg$^2$ \citep{GWTC2}.  Nonetheless, associations with specific host galaxies are impossible for all observed binary black hole mergers.  

\textbf{Eccentricity:}  GW190521 appears to have residual eccentricity of at least 0.1 at the gravitational-wave frequency of 10 Hz \citep{RomeroShaw:2020GW190521}.  One or more other binary black hole mergers may also show signs of non-negligible eccentricity \citep{RomeroShaw:2021}.  However, there is a possible degeneracy between eccentricity and precession \citep{Bustillo:2021}, and analyses generally do not include both effects, so eccentricity observations do not appear conclusive at this time.

\textbf{Testing General Relativity:} All gravitational-wave signals observed to date are consistent with gravitational waves expected within the general theory of relativity, providing a stringent test of this theory in the dynamical, strong-field regime \citep{GW150914:GR,GWTC2:GR}.

\textbf{Merger Rate:} The uncertainty in the binary black hole mass distribution is relayed into the uncertainty on the inferred merger rate, [16,130]  Gpc$^{-3}$ yr$^{-1}$ for the combined interval 90\% credible interval across mass distribution models under the assumption of a constant-in-redshift merger rate.  However, there is strong evidence that the merger rate increases with redshift; the local merger rate at redshift $z=0$ is [10,30]  Gpc$^{-3}$ yr$^{-1}$  \citep{GWTC3:pop}.  

\subsection{Electromagnetic observations}

While recent gravitational-wave observations have invigorated the field of massive binary evolution, there already exists a wealth of electromagnetic observations of systems at various stages along the possible paths to binary black hole formation. In this section we describe some of the key electromagnetic observations that shed light on compact object binary formation and evolution preceding the merger.

A wide variety of electromagnetic observations inform our understanding of the evolution of massive stellar binaries. These include observations of: 
\begin{itemize}
\item the initial mass and period distributions of binary stars at formation \citep[e.g.,][]{Sana:2012,MoeDiStefano:2017}; 
\item luminous red novae, which may be associated with common envelope events \citep[e.g.,][]{Ivanova:2013LRN,Howitt:2020};
\item Galactic binary radio pulsars \citep[e.g.,][]{Tauris:2017};
\item short gamma-ray bursts \citep[e.g.,][]{Berger:2014};
\item supernovae and long gamma-ray bursts \citep[e.g.,][]{Cantiello:2007,Szecsi:2017,Bavera:2021};
\item X-ray binaries \citep[e.g.,][]{TaurisvdH:2006,Fabbiano:2006};
\item black holes in detached binaries \citep[e.g.,][]{Thompson:2019} and microlensing observations \citep[e.g.,][]{WyrzykowskiMandel:2019}.
\end{itemize}
We first focus on X-ray binaries, the type of binary system with the closest connection to merging binary black holes, although only a small minority of black-hole X-ray binaries will ultimately evolve into merging black holes \citep[e.g.,][]{CygnusX3:2012,Neijssel:2020CygX1}. We then discuss detached black-hole binaries, black-hole microlensing observations, and selection effects.

\subsubsection{X-ray binaries}\label{sec:XRB}

Black-hole X-ray binaries consist of a star transferring mass onto a black-hole companion, leading to the emission of X-ray radiation from the accretion disk surrounding the black hole.  X-ray binaries containing accreting black holes can be divided into two categories, low-mass and high-mass, in reference to the mass of the black hole's companion star. The accretion process differs between the two types: in low-mass X-ray binaries, the tidal gravitational pull of the black hole causes mass to stream from the companion onto the black hole in the process known as Roche-lobe overflow, while in high-mass X-ray binaries, the black hole accretes only a fraction of the material (stellar wind) driven off the companion's surface.

\textbf{Black-Hole Masses:} 
Until recently, black-hole X-ray binaries with dynamical mass measurements provided the only secure measurements of black-hole masses other than the gravitational-wave observations described above, though this is now changing thanks to observations of detached binaries and microlensing data (see section \ref{sec:lensing}).  \citet{Ozel:2010} and \citet{Farr:2011} summarise the mass distribution of black-hole X-ray binaries.  They find a distribution which ranges from $> 20\, M_\odot$ for the heaviest high-mass X-ray binaries to 4 or 5 solar masses for the lightest low-mass X-ray binaries, with a possible mass gap between neutron-star and black-hole masses, though see \cite{Kreidberg:2012} for possible biases in mass measurements of some low-mass X-ray binaries and section \ref{sec:lensing} for evolutionary selection effects.  The uncertainty regarding the radial velocity measurements in high-mass X-ray binaries with Wolf-Rayet companions such as IC10 X-1 and NGC 300 X-1 \citep{Laycock:2015} means that the most massive confirmed stellar-mass black hole is Cygnus X-1, with a dynamical mass measurement of $21 \pm 2\, M_\odot$ \citep{MillerJones:2021}.   The measured masses of black holes in X-ray binaries are sketched in figure \ref{fig:BHmasses}, along with the masses of the observed gravitational-wave sources.  This figure does not include the speculative but potentially very exciting evidence for intermediate-mass black holes \citep{MillerColbert:2004,Greene:2019,Paynter:2021}; if few-hundred solar-mass black holes generically exist in globular clusters, they will substitute into merging binaries and ultimately be observable as gravitational-wave sources \citep[e.g.,][]{Mandel:2008,IMBBH:O1}.

\textbf{Black-Hole Spin Magnitudes:} Black-hole X-ray binaries also provide an opportunity for measuring black-hole spins \citep[see][for recent reviews]{MillerMiller:2015,Reynolds:2020}.  Continuum fitting of the X-ray flux from the accretion disk and iron K-$\alpha$ line fits to the disk reflection profile can both be used to infer the location of the inner edge of the disk, which is assumed to correspond to the radius of the innermost stable circular orbit, a sensitive function of black-hole spin.  Quasi-periodic oscillations also have the potential to provide spin measurements.  Unfortunately, the underlying physical mechanisms are not fully understood at present, and the inferred spins may suffer from significant systematics \citep[e.g.,][]{Basak:2017,Kawano:2017}.  

\textbf{Black-Hole Spin Directions:} We know even less about the spin directions than the spin magnitudes.  Although black-hole spin-orbit alignment is assumed in continuum flux measurements \citep{MillerMiller:2015}, some black-hole X-ray binaries, including GRO J1655-40 \citep{Martin:2008}, 4U 1543-47 \citep{MorningstarMiller:2014} and particularly V4641 Sgr \citep{Orosz:2001,Martin:2008b} appear to indicate that the microquasar jet, presumably aligned with the BH spin axis, is misaligned with the orbit.  Moreover, initial stellar spins in binaries have been observed to be misaligned \citep[e.g.,][]{Albrecht:2009,Albrecht:2014}, though there are opportunities for subsequent realignment during binary evolution, as described below.

\textbf{Inconsistency with Spin Observations from Gravitational-wave Sources?} 
All three high-mass X-ray binaries with available spin measurements -- Cygnus X-1, LMC X-1 and M33 X-7 -- show evidence of rapid dimensionless spins in excess of $\gtrsim 0.85$, where 1 is the spin magnitude of a maximally spinning black hole \citep{Reynolds:2020}.  This is significant, because unlike long-lived low-mass X-ray binaries, whose spin magnitudes could be altered by accretion from the companion, especially if they started out with intermediate-mass companions \citep{Podsiadlowski:2003,Fragos:2015}, high-mass X-ray binaries are too short-lived to enable significant spin changes due to accretion \citep{KingKolb:1999}.  Isolated binaries that form binary black holes are expected to go through the high-mass X-ray binary phase during their evolution (though \citet{HiraiMandel:2021} argue that only binaries with optical companions that are close to Roche-lobe overflow may be X-ray bright).  Thus, if the high spin magnitude measurements in black-hole X-ray binaries are to be believed, then the observed high-mass X-ray binaries and merging binary black holes should predominantly sample different evolutionary histories (see section \ref{BHspins}).

\subsubsection{Other electromagnetic black hole mass measurements} \label{sec:lensing}

\textbf{Detached binaries:} Until recently, only stellar-mass black holes in X-ray binaries allowed accurate dynamical mass measurements.  However, this has recently changed with observations of wide binaries with a luminous companion detached from a non-accreting black hole.  Examples of such observations include two $\sim 3\, M_\odot$ black holes in detached binaries \citep{Thompson:2019,Jayasinghe:2021} along with several black holes in detached binaries in globular clusters \citep{Giesers:2019,Saracino:2021}.  Surveys like Gaia may detect hundreds of detached black hole binaries \citep[e.g.,][]{Chawla:2021}.  However, the case of LB-1, which was initially thought to contain an extremely massive black hole in a wide binary, highlights some of the challenges of accurately measuring black hole masses in detached systems \citep{Liu:2019,Eldridge:2020,AbdulMassih:2020,Shenar:2020}.   

\textbf{Microlensing:} Microlensing observations provide a tool for measuring the masses of single black holes.  When such black holes pass between a luminous star and the observer, they temporarily lens the star, increasing its luminosity while the alignment is roughly within the lens's Einstein ring.  Mircolensing measurements, together with gravitational-wave and detached black-hole binary observations, suggest that there is no evidence for a mass gap between neutron stars and black holes \citep{WyrzykowskiMandel:2019, Mroz:2021}.  However, there is a possible degeneracy between lower lens masses and higher lens-source relative velocities, since both reduce the transient duration; thus, the low-mass microlensing black-hole candidates could be consistent with higher-mass black holes receiving natal kicks of $\gtrsim 80$ km s$^{-1}$.

\textbf{Selection effects:} Finally, it is worth remembering that all observations are likely to suffer from a range of selection effects.  These could be Malmquist biases which favour observations of the brightest or loudest sources, such as the enhanced detectability of more massive binaries in gravitational-wave data or of more rapidly accreting X-ray binaries (though more complex detectability thresholds related to wind morphology may be relevant for high-mass X-ray binaries \citep{HiraiMandel:2021}).  Alternatively, they could be evolutionary selection effects which can skew the observed populations; for example, X-ray binaries may contain few low-mass black holes, creating an impression of a mass gap between neutron star and black hole masses, if low-mass black holes get larger natal kicks that tend to disrupt binaries \citep{Mandel:2020}.

\section{Formation scenarios}\label{form}

When the detection of GW150914 was first announced, many were surprised that it was a binary black hole rather than a binary neutron star: there was already observational evidence for merging double neutron stars in the Galaxy (starting with the Hulse-Taylor binary pulsar \citep{HulseTaylor:1975}), but there was no direct evidence for merging binary black holes. Some were even more surprised by the high black-hole masses, in excess of the observed black-hole masses in X-ray binaries. Was this surprise justified? And should we be perplexed by the spin or mass ratio measurements for the gravitational-wave sources? Before answering these questions, we should first discuss how these merging compact binaries come into existence. In fact, this is one of the key outstanding questions that gravitational-wave observations will help us to answer. Below, we explain why -- on the face of it -- it is rather startling that compact massive binaries exist at all. We then describe the leading candidate formation channels, and in the next section we discuss the ways in which gravitational-wave observations might be used to distinguish them.

\subsection{The Orbital Separation Question}

\begin{figure}
	\centering
	\includegraphics[width=0.8\textwidth]{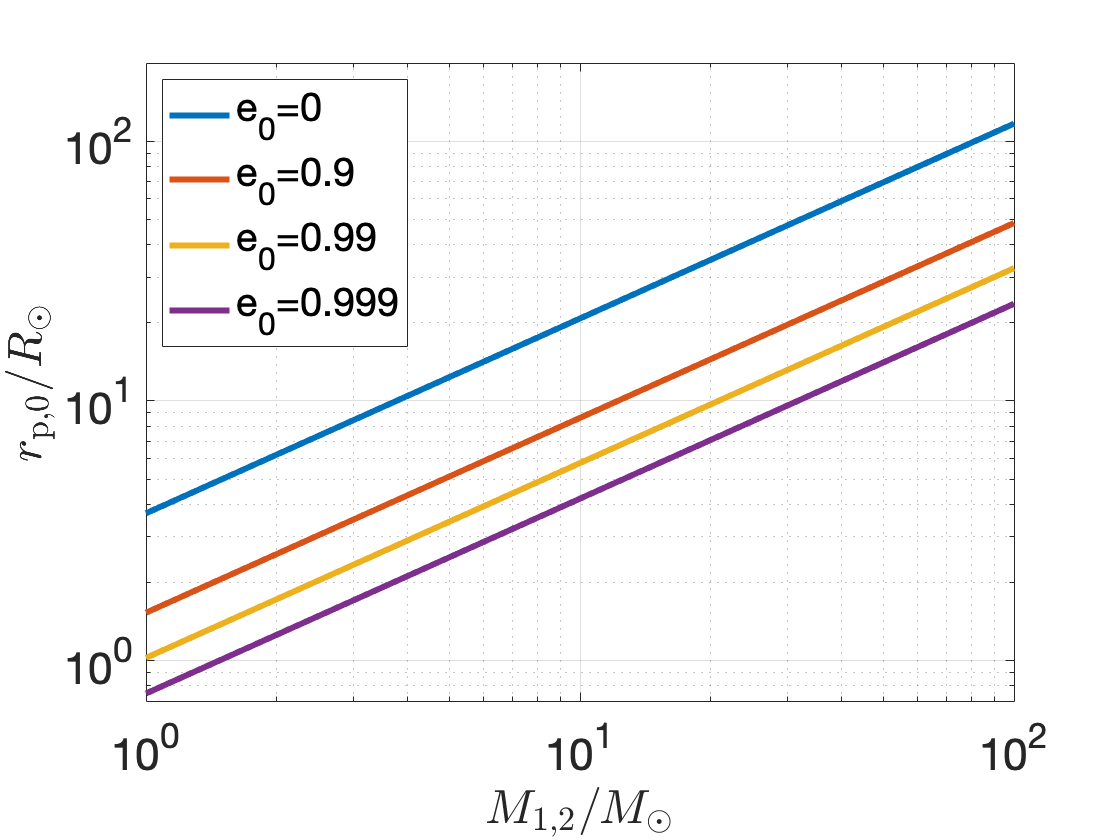}
	\caption{Maximum initial periapsis separation (ordinate) that a binary black hole with equal-mass components (abscissa) can have while still merging within 14 Gyr through gravitational-wave emission, for four different choices of initial eccentricity: $e=0$, 0.9, 0.99, and 0.999 from the top down.  Computed using the \citet{Mandel:2021} fit to \citet{Peters:1964}. \label{fig:periapsis}}
\end{figure}

\textbf{Only very tight binaries can merge via gravitational waves.} Gravitational-wave emission is a very strong function of orbital separation. During a compact binary merger, the luminosity in gravitational waves is a few thousandths of the Planck luminosity, $c^5/G$ \citep[e.g.,][]{Cardoso:2018}; at nearly $10^{57}$ ergs per second, such mergers ``outshine'' all the stars in the visible Universe combined. But because gravitational-wave luminosity is inversely proportional to the fifth power of the binary separation \citep{Peters:1964}, widely separated binaries lose energy very slowly and undergo only negligible inspiral over billions of years. Only very close binaries can be brought to merger by gravitational waves within the age of the universe; \autoref{fig:periapsis} shows the maximum initial separation for an equal-mass binary to merge on this timescale. For the two $\sim 30 M_\odot$ black holes responsible for GW150914, the initial separation at periapsis must have been less than $\sim 50 R_\odot$ -- just a quarter of the distance from the Earth to the Sun -- if the merger was driven by gravitational-wave emission alone.

\begin{figure}
	\centering	
	\includegraphics[width=0.8\textwidth]{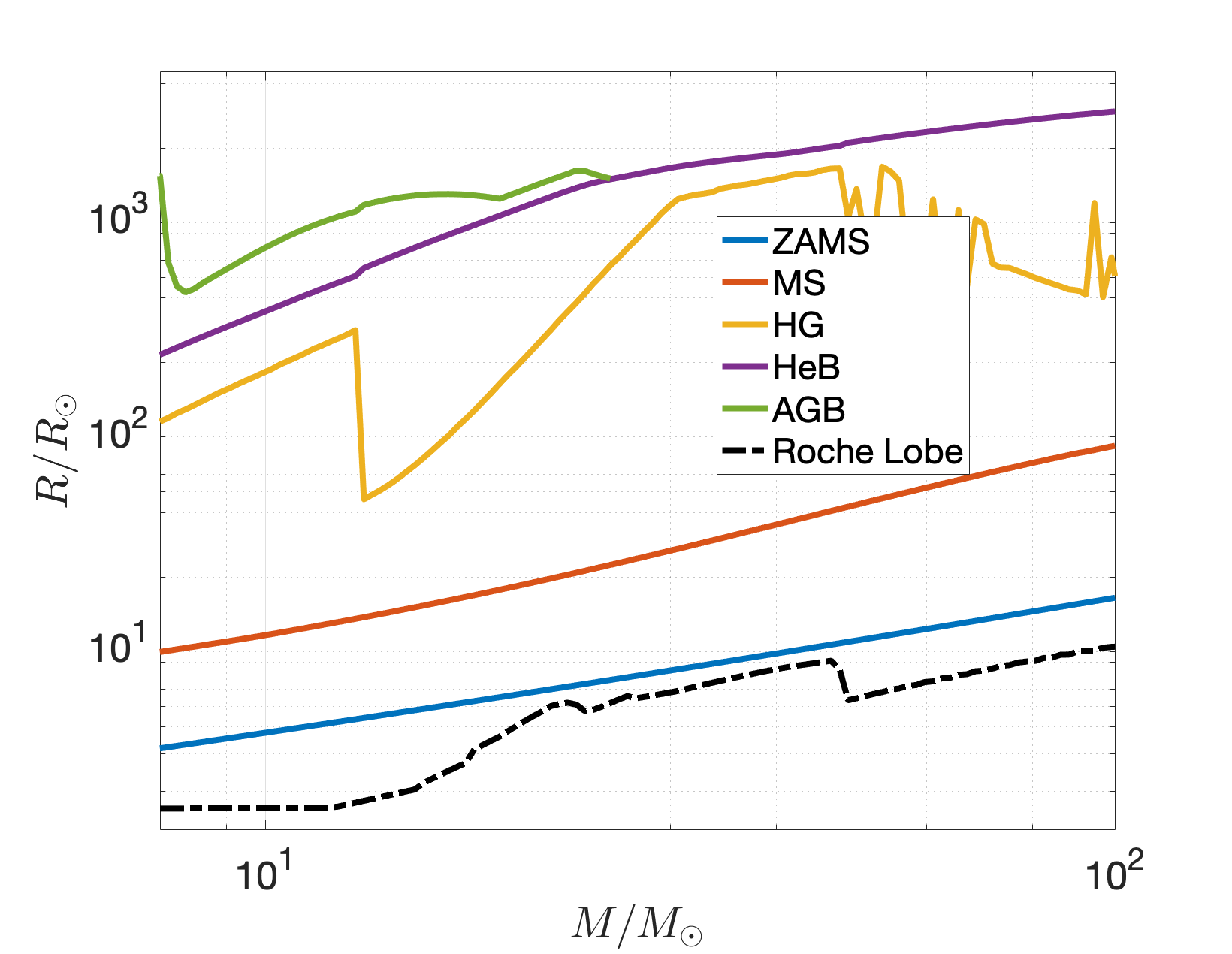}
\caption{Maximal stellar extent at solar metallicity during various phases of stellar evolution for a non-rotating star with a given initial mass; based on the implementation of single stellar evolution models of \citet{Hurley:2000} in the COMPAS binary population synthesis code \citep{COMPAS:2021}.   The star expands from its zero-age main sequence (ZAMS) size onwards through the main sequence (MS), Hertzsprung gap (HG), core Helium burning (HeB) stages, and the asymptotic giant branch (AGB -- here also used colloquially for hydrogen-rich massive stars with a carbon/oxygen core).  The Roche lobe radius (maximal stellar size beyond which a star would engage in mass transfer \citep{Eggleton:1983}) is plotted for an equal mass circular binary which would have the maximum separation to allow a merger within the age of the Universe (see \autoref{fig:periapsis}) assuming initial mass to compact object mass conversion as in \autoref{fig:BHremnant}.\label{fig:Rmax} }
\end{figure}

\textbf{But the black holes' parent stars cannot get so close.} Stars expand as they evolve. Even our Sun will reach roughly an astronomical unit in size during its giant phase; the stars that leave behind black holes (above $\sim 20 M_\odot$  at birth) may reach thousands of solar radii at their maximal extent. The maximum stellar radius as a function of initial mass is plotted in \autoref{fig:Rmax}. There appears to be a problem. If the parent stars begin life in a close binary with a separation from which gravitational waves could bring their remnants together, the stars will expand to sizes much larger than their separation as they evolve, and we might therefore expect that they would merge long before they collapse into black holes. If they start sufficiently far apart to avoid merger before collapse into black holes, their remnant binaries will take many millions of times longer than the age of the Universe to merge. In either case, no gravitational-wave sources would exist today. 

\textbf{The problem is already there at birth.} In fact, for reasonable models of wind-driven mass loss and mass loss during supernovae, even the initial stellar radii at the start of the main sequence are too large to fit into a binary that could merge within the age of the Universe just through gravitational-wave emission.  The black `Roche Lobe' curve in \autoref{fig:Rmax} shows the maximum size that a star could have in order to fit into a circular, merging equal-mass binary (see \autoref{fig:periapsis}).  This size is smaller than the zero-age main sequence radius for all stellar masses shown in the figure: the separation problem thus exists even without accounting for the exacerbating factors of stellar expansion or binary widening through mass loss.

\textbf{So then why do the gravitational-wave sources exist at all?} Given the above, one might conclude that gravitational-wave driven compact binary mergers should not exist. Yet their existence and detection were expected \citep{ratesdoc}. In fact, \citet{Dyson:1962} conjectured about the existence of merging neutron star binaries even before the first neutron star was observed; \citet{Tutukov:1973} predicted that binary compact objects must naturally (albeit rarely, and at wide separations in their model) form as a result of massive binary evolution; \citet{vdHDeLoore:1973} argued that tight high-mass X-ray binaries -- progenitors of compact object binaries -- must also form; and the Hulse--Taylor binary pulsar \citep{HulseTaylor:1975} demonstrated the existence of binary compact objects that would merge within the age of the Universe \citep[for early explanations of its formation in the context of binary evolution, see][]{FlanneryvdH:1975,DeLoore:1975}.

The proposed formation scenarios for merging compact-object binaries generally fall into two broad categories: the evolution of isolated binaries composed of two massive stars and binaries which form or merge with the assistance of gravitational dynamical interactions.  Below, we discuss four examples of these scenarios: (i) finely tuned binary evolution that brings the stars closer as they expand and interact; (ii) finely tuned stellar evolution that prevents the parent stars from expanding at all; (iii) assembly of a close binary from black holes that formed from stars not born in the same binary; and (iv) stellar and binary evolution and dynamics operating jointly in triple systems.   Contrived as these scenarios sound, all four might plausibly contribute to the production of binary black hole mergers.  And as we will see in \autoref{merge}, only a small fraction of stars, of order one in a million, need to end up in merging black-hole binaries in order to explain the observed merger rates: even quite unusual or (apparently) finely tuned evolutionary pathways could therefore be viable candidates.

We explore these four formation scenarios in more detail below. We will not discuss some of the more exotic proposed mechanisms, such as the fragmentation of a single stellar core into two black holes \citep{Loeb:2016} (\citet{Woosley:2016} and \citet{Dai:2017} discuss the problems with this picture); cosmological coupling \citep{Croker:2021}; or physics beyond the standard model \citep{Sakstein:2020}.   

We limit our discussion to black holes of astrophysical origin.   Mergers of primordial black holes formed through the collapse of early-Universe density perturbations have been proposed as sources of gravitational waves \citep{Bird:2016}.   The contribution of primordial black hole mergers to the observed population of gravitational-wave signals is sensitive to the masses of primordial black holes, the total fraction of dark matter they comprise, their initial distribution in binaries or small clusters, and the possibility of gas accretion onto primordial black holes \citep{AliHaimoud:2017, ChenHuang:2018, Korol:2019, DeLuca:2020b, DeLuca:2020a}.

\subsection{The Candidate Formation Scenarios}
\subsubsection{Coming closer later in life: classical isolated binary evolution via the common-envelope phase}
\label{form:isol}

The first channel we describe is perhaps the most studied one, placing merging black-hole binaries in the same framework as other very close binaries (such as cataclysmic variables) in which at least one of the stars has extended beyond the binary's current orbital separation during an earlier stage of its evolution \citep[e.g.,][]{Paczynski:1976}. 

In this scenario the two stars are born in a relatively wide binary, allowing them space to expand.  However, at a critical moment in its evolution, the binary is tightened by a factor of two or more orders of magnitude through dynamically unstable mass transfer, known as a common envelope phase \citep{LivioSoker:1988,Ivanova:2013}. The resulting tight binary may then be close enough to merge through gravitational-wave emission.  \citet{SmarrBlandford:1976} may have been the first to explicitly point to this channel in the context of compact object binary formation when analysing the evolutionary history of the Hulse--Taylor binary pulsar. The channel has been studied at length over the past 40 years, with significant contributions from \citet{TutukovYungelson:1993,Lipunov:1997,BetheBrown:1998,Nelemans:2001,Belczynski:2002,VossTauris:2003,Pfahl:2005,Dewi:2006,Kalogera:2007,OShaughnessy:2008,Dominik:2012,Mennekens:2014,Belczynski:2016,EldridgeStanway:2016} and many others. Rather than summarising all of the steps and challenges in our understanding of massive binary evolution (see the papers above and the review by \citet{PostnovYungelson:2014} for details), we provide a schematic outline of an evolutionary scenario that leads to the formation of a GW150914-like merging system.

\begin{figure}
	\centering
	\includegraphics[width=0.45\textwidth]{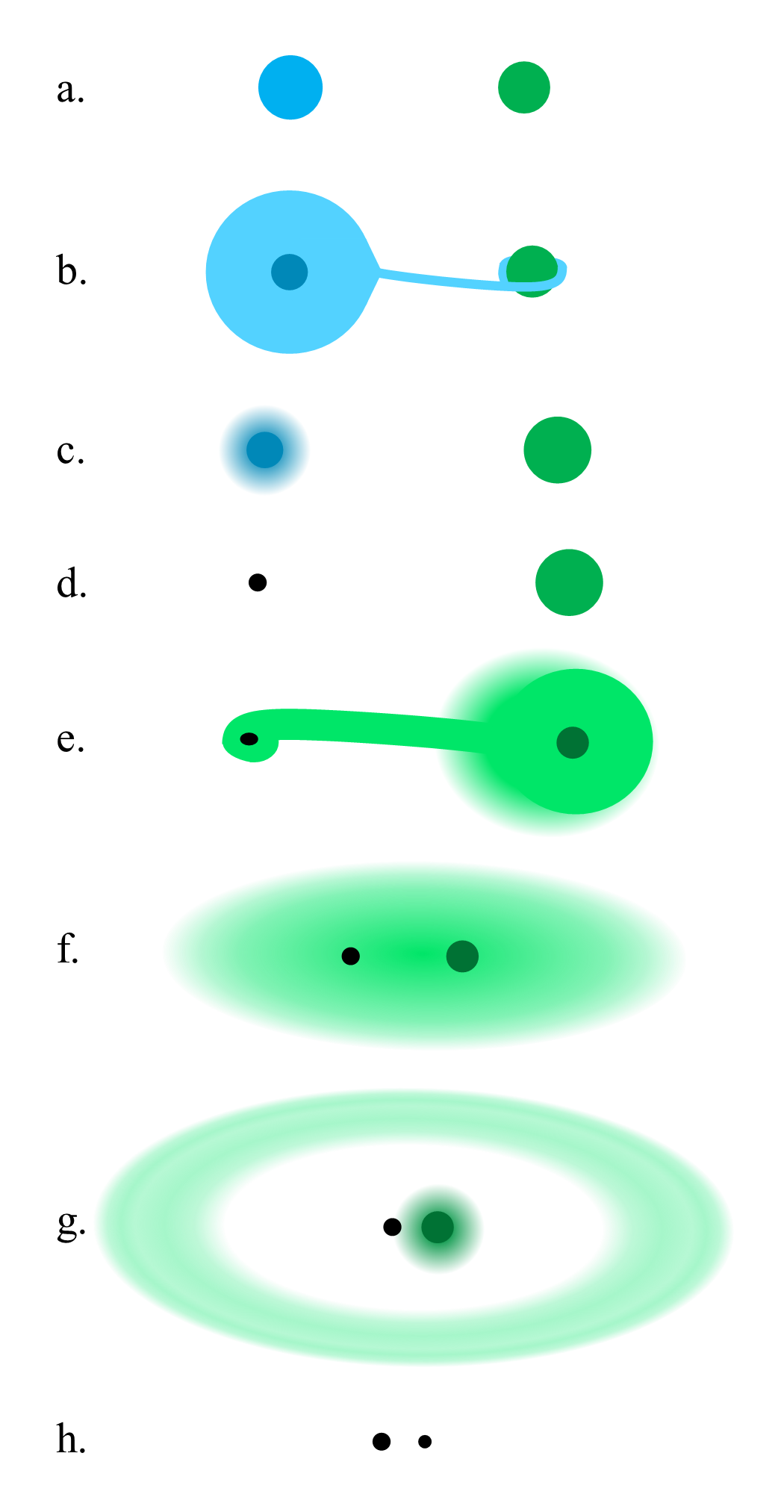}
	\caption{\label{fig:isol_binary} A sketch of merging black-hole binary formation through isolated binary evolution via the common envelope phase.  The steps are described in section \ref{form:isol}.}
\end{figure}

An example of the evolution of this system is sketched out in a van den Heuvel -- style diagram in \autoref{fig:isol_binary}, and the steps below follow the panels in that figure:
\begin{enumerate}  
\item[a.] Two massive stars of perhaps $100$ and $75$ $M_{\odot}$ are born in a binary at a separation of $\sim 10$ AU in a low-metallicity environment ($\sim 5\%$ of solar metallicity, or metal fraction --- the fraction $Z$ of the star's mass contained in elements heavier than hydrogen and helium).  
\item[b.] The more massive primary reaches the end of its main sequence evolution first.  At this stage, it has completed fusing hydrogen into helium in its core, and with the loss of energy input, the core begins to contract.  The associated release of gravitational binding energy and the eventual onset of hydrogen shell burning cause the hydrogen-rich envelope to expand.  For a sufficiently close binary, the primary expands past the equipotential surface known as the Roche Lobe and begins to transfer mass onto the secondary.  This mass transfer proceeds on the thermal timescale of the primary donor and could be significantly non-conservative, as the less evolved secondary, with its correspondingly longer thermal timescale, is unable to accept mass at the rate at which it is being donated.  The loss of mass from the binary, in addition to wind-driven mass loss, can widen the system to perhaps $\sim 20$ AU.  
\item[c.] The primary loses its entire envelope, leaving behind a naked helium-burning star -- a Wolf-Rayet star.  
\item[d.] Following wind-driven mass loss, which further widens the system, the primary collapses into a black hole; this collapse may be complete, or there may be some ejected mass and an associated natal kick.  
\item[e.] When, a few hundred thousand years later, the secondary reaches the end of its main sequence, the process repeats in reverse: the secondary expands until it commences mass transfer onto the primary.  By this time, the primary has lost around two thirds of its initial mass through a combination of envelope stripping, winds, and possible mass loss during a supernova (if the fallback is not complete).  The mass transfer on to the black hole would need to be almost wholly non-conservative if the accretion obeys the Eddington limit, which corresponds to an equilibrium between gravity and the pressure of the radiation released during accretion on infalling material.  Mass transfer from a massive donor to a lower-mass accretor adds specific angular momentum to the transferred mass, so conservation of angular momentum requires that binary's orbit to shrink.  Consequently, mass transfer would lead to a rapid hardening of the binary at a rate that is faster than the reduction in the size of the secondary donor as it loses mass.  As a result, the more mass it donates, the more the donor overflows its Roche lobe.  
\item[f.] This runaway process of dynamically unstable mass transfer leads to the formation of a common envelope of gas (from the donor's envelope) around the binary.  The drag force on the black hole from the envelope leads to rapid spiral-in.   The dissipated orbital energy is deposited in the envelope, and may ultimately lead to the expulsion of the envelope.  
\item[g.] After the expulsion of the common envelope, the resulting black hole -- Wolf-Rayet binary has a separation of only $\sim 35 R_\odot$ in this example.  
\item[h.] Following further wind-driven mass loss from the secondary and its collapse into a black hole, the black-hole binary is formed.  While this entire process takes only a few million years from the formation of a stellar binary to the formation of a binary black hole, the subsequent inspiral through gravitational-wave emission will last for around 10 billion years before merger.
\end{enumerate}

Of course, this relatively simple picture holds many uncertainties: the rate of mass loss through winds, particularly during specific stellar evolutionary phases such as from luminous blue variables (massive supergiant stars with significant outbursts and eruptions and associated rapid mass loss \citep{Vink:2011})  and its dependence on metallicity; the fraction of the donated mass that is added to the accretor during stable mass transfer \citep{KippenhahnMeyerHofmeister:1977} and the specific angular momentum of the mass that is removed from the binary; the response of a star to mass loss and the onset of a common envelope phase \citep{Pavlovskii:2017}; common-envelope survival and the amount of binary hardening associated with the envelope ejection \citep[e.g.,][]{Kruckow:2016,Klencki:2020convective,Fragos:2019,LawSmith:2020,Lau:2021}; the amount of mass that may be retained by a star after its envelope is stripped off \citep[e.g.,][]{Laplace:2020}; supernova fallback and natal kicks for black holes \citep[e.g.,][]{Fryer:2012,Repetto:2012,Mandel:2015kicks,Mueller:2020}; the possibility of double-core common envelopes when unstable mass transfer is initiated between two evolved stars \citep{BetheBrown:1998,Belczynski:2002,Dewi:2006}; and the amount of mass that can be accreted onto a black hole \citep{Eldridge:2017,Bavera:2020,vanSon:2020}.  Moreover, some isolated binaries could form merging binary black holes entirely through dynamically stable mass transfer, with sufficient binary hardening achieved by non-conservative mass transfer onto a lower-mass compact donor rather than a common-envelope event \citep{vandenHeuvel:2017,Pavlovskii:2017,Neijssel:2018,vanSon:2021}.  We will discuss the possibilities of addressing some of these with future gravitational-wave observations in \autoref{prospect}.

\subsubsection{Abracadabra, thou shalt not expand: chemically homogeneous evolution}\label{form:CHE}

\begin{figure}
	\centering
	\includegraphics[width=0.4\textwidth]{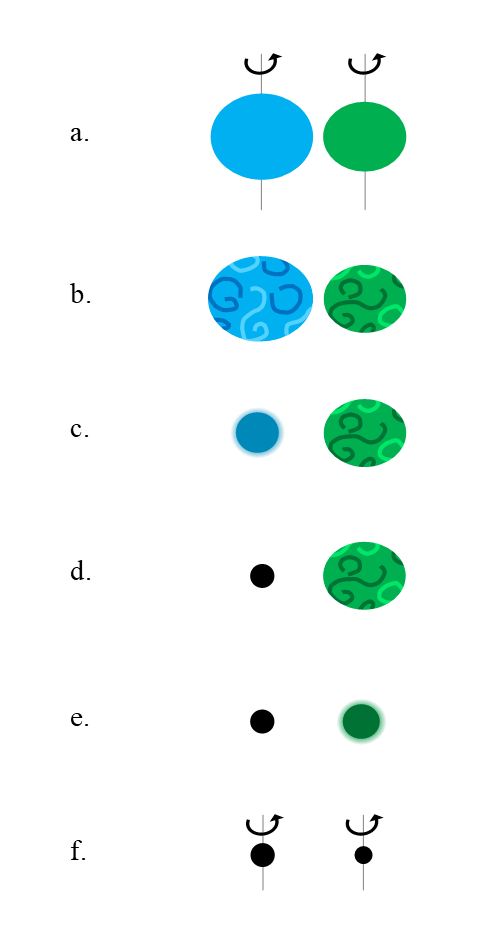}
	\caption{\label{fig:chem_homog} A sketch of merging black-hole binary formation via the chemically homogeneous evolution channel.  The steps are described in section \ref{form:CHE}.}
\end{figure}

What if some types of massive stars did not expand and converted most of their mass into a black hole? A massive binary could then start out at an orbital period of a couple of days, close enough that if the stars produced black holes in situ they would merge within the age of the Universe through gravitational-wave emission. Without an expansion phase, the parent stars would be in no danger of merger; if almost all of the star's mass was converted into a black hole rather than the $\sim$30--50\% of the mass which typically becomes the helium core (see \autoref{fig:BHremnant}), the merger timescale would shrink for a given separation, lifting the Roche lobe radius curve in \autoref{fig:Rmax} and removing the ``separation problem at birth''.  No fine-tuning of binary evolution would then be required to construct a plausible formation scenario for tight black-hole binaries.   Efficient mixing in high-mass, low-metallicity stars in close binaries, known as chemically homogeneous evolution, could enable this scenario.

This evolutionary pathway is sketched out in \autoref{fig:chem_homog}:
\begin{itemize}
\item[a.] Binary companions raise tides on each other, much like the Moon's tides on Earth.  If a binary is tight enough that each star fills a significant fraction of its Roche lobe, tidal energy dissipation is rapid and proceeds until the stars are tidally locked, i.e. the rotation periods of the stars are synchronised to the orbital period of the binary. This also means that the stars are rotating at a few tens of percent of their break-up velocities.  
\item[b.] Such rapidly rotating stars will develop significant temperature gradients between the poles and the equator, which may lead to efficient large-scale meridional circulation within each star \citep{Eddington:1925,Sweet:1950}.  \citet{EndalSofia:1978} and subsequent studies \citep[e.g.,][]{Heger:2000,MaederMeynet:2000,Yoon:2006,Szecsi:2015} explored the internal shears and their impact on the mixing of chemical species within the star.  Although quantitative predictions differ, it appears that rapidly rotating massive stars may efficiently transport hydrogen into the core and helium out into the envelope until nearly all of the hydrogen in the star is fused into helium.  
\item[c--f.] Then, at the end of the main sequence, the star behaves essentially as a Wolf-Rayet naked helium star, contracting rather than expanding. As long as the metallicity is sufficiently low that the wind-driven mass loss does not significantly widen the binary -- which would lead to the loss of co-rotation and chemically homogeneous evolution \citep{deMink:2009} -- the binary can avoid mass transfer.  
\end{itemize}

\citet{MandeldeMink:2016,Marchant:2016,deMinkMandel:2016} explored such binaries and concluded that they could present a viable channel for forming the most massive observed gravitational-wave sources (such as GW150914), though not the lowest-mass systems (see also \cite{duBoisson:2020,Riley:2020}).  This scenario does, however, require some unproven assumptions regarding stellar evolution, with uncertain or even contradictory observational supporting evidence \citep{Almeida:2015, AbdulMasih:2021}.

\textbf{Population III stars:} A somewhat similar evolutionary channel to chemically homogeneous evolution --- one in which stars in a close binary avoid significant radial expansion --- could play out for the first generation of metal-free stars in the early Universe, known as population III stars.  The absence of primordial carbon, nitrogen and oxygen in these stars could prevent the CNO cycle hydrogen fusion in the hydrogen shell around the helium-rich core, reducing radial expansion after the end of the main sequence \citep{Marigo:2001}.  In addition to removing the need for subsequent mass transfer as in the chemically homogeneous evolution channel, this could allow the star to retain its tightly bound hydrogen envelope when collapsing into a black hole \citep{Kinugawa:2021}.  Population III stars would also experience reduced mass loss through winds due to their zero metallicity.   Meanwhile, the initial mass function of zero-metallicity stars could itself favour higher masses.  All of these effects make population III stars promising progenitors for the formation of merging binary black holes \citep{Belczynski:2004popIII,Kinugawa:2014,Inayoshi:2017}.  However, there are significant uncertainties in the initial properties of population III binaries and the amount of stellar expansion that population III stars will experience \citep{Hirano:2014,Stacy:2016,Hartwig:2016,Belczynski:2017popIII}.

\subsubsection{The black-hole matchmaking club: dynamical formation in dense stellar environments}\label{form:dyn}

\begin{figure}
	\centering
	\includegraphics[width=0.4\textwidth]{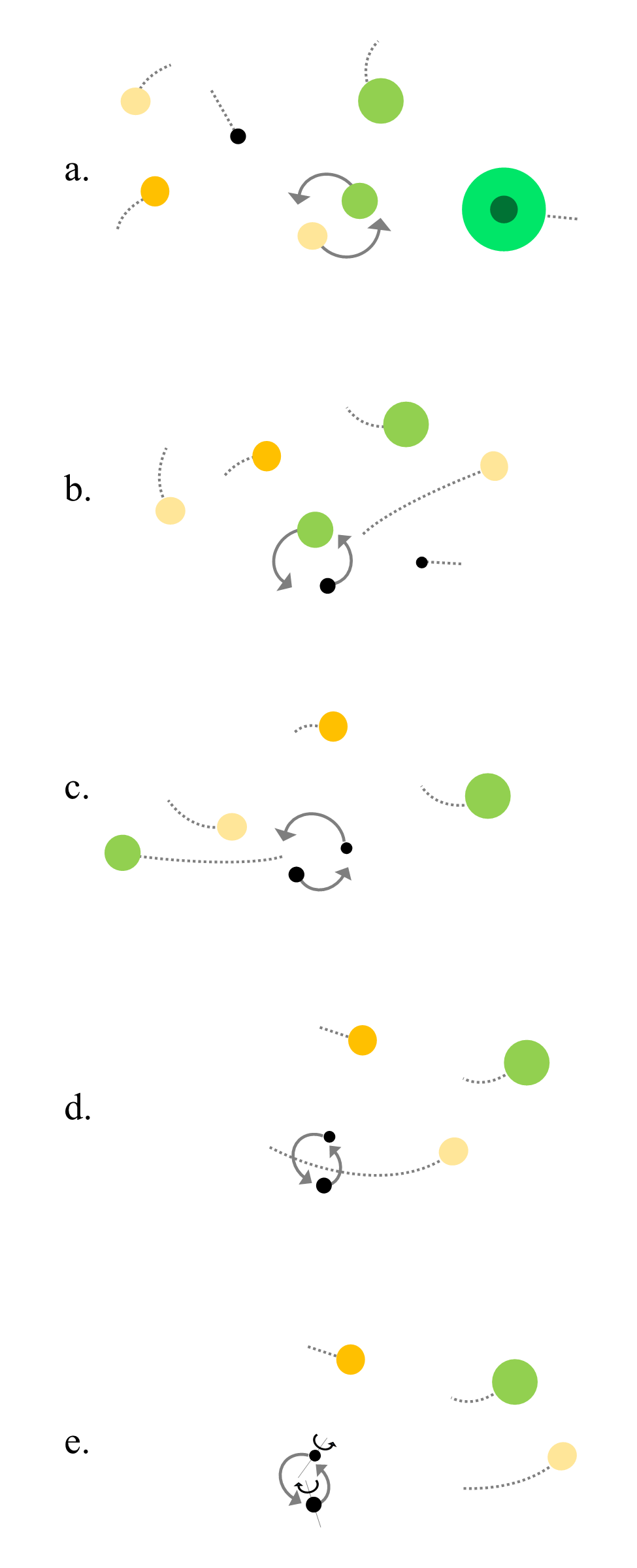}
	\caption{\label{fig:dynamical} A sketch of merging black-hole binary formation via the dynamical evolution channel.  The steps are described in section \ref{form:dyn}.}
\end{figure}

The merging black holes may not have formed in the same binary at all.  Instead, in this channel, they are remnants of independent massive stars (perhaps single, or each in its own separate binary), but are then introduced to each other by a matchmaker, or rather a whole array of matchmakers.   This evolution is illustrated in \autoref{fig:dynamical}: 
\begin{enumerate}
\item[a.] Suppose the two black holes formed in a dense stellar environment such as a globular cluster.  
\item[b.] As the most massive objects in the cluster, the black holes mass segregate toward the cluster centre as kinetic energy is equipartitioned by dynamical scattering interactions \citep{Spitzer:1969,BinneyTremaine}, but see \citep{Trenti:2013}.  Once there, they may form binaries through three-body interactions, in which one of three initially unbound objects is endowed with sufficient kinetic energy to leave a bound binary, or by substituting into existing binaries: in a $2+1$ interaction, the lightest object is usually ejected in favour of the two heavier objects forming a binary \citep{HillsFullerton:1980}.  
\item[c,d.] If the orbital speed of the binary's components is greater than the typical speed of stars in the cluster (the binary is `hard'), then subsequent interactions with other objects in the cluster -- the matchmakers -- will gradually tighten the black-hole binary \citep{Heggie:1975}. The interlopers are likely to leave with slightly higher speeds than they arrived with, each time removing energy from the binary.   Meanwhile, `soft' binaries, with orbital speeds smaller than the velocity dispersion in the cluster, will be disrupted. 
\item[e.] If the density of objects is high enough to ensure a suitable rate of interactions, the binary will be hardened until it is compact enough to merge through gravitational-wave emission, provided it does not get kicked out of the cluster through a recoil kick from the last interaction -- and even then, it may still go on to merge outside of the cluster.  
\end{enumerate}

The prospects for these dynamically formed merging binary black holes in globular clusters and young massive stellar clusters have been explored through a series of analytical estimates and numerical experiments by \cite{Sigurdsson:1993,Kulkarni:1993,PZwart:2000,OLeary:2006,Banerjee:2010,Downing:2011,Morscher:2015,Rodriguez:2016,Askar:2016,Park:2017,FragioneKocsis:2018,Mapelli:2020} and others.  \citet{OLeary:2008} \citep[but see][]{Tsang:2013} and \citet{AntoniniPerets:2012} explored dynamical formation in galactic nuclei with a massive black hole (see section \ref{ecc} for a discussion of capture through gravitation-wave emission), while \citet{MillerLauburg:2008} and \citet{AntoniniRasio:2016} argued that galactic nuclei without a massive black hole are a favoured site for dynamical formation.  

An accretion disk in an active galactic nucleus can act as a catalyst for binary black hole mergers \citep{Bellovary:2016, Bartos:2016, Stone:2016, McKernan:2018, Tagawa:2020}.   Compact objects in active galactic nuclei may form directly in the disk, be captured by the disk through gas drag, or settle into a disk through resonant relaxation against the stellar background \citep{SzolgyenKocsis:2018}.  Once in the disk, black holes can be transported radially until, perhaps, being captured into migration traps, where high black-hole concentrations increase interaction rates \citep{Bellovary:2016,Yang:2019,Secunda:2019,PengChen:2021}.  Once binaries form they can be hardened by a combination of dynamical interactions and gas accretion.  As with other mergers in galactic nuclei, the high escape velocity makes it more likely that merger products will be retained despite gravitational-wave recoil kicks, enabling hierarchical mergers \citep{Yang:2019,Secunda:2020,Tagawa:2021}.  While active galactic nuclei are promising environments for black-hole binary formation, some of the many uncertainties include star formation in disks, the efficiency of migration traps \citep{PanYang:2021}, the details of gas interaction with black holes and binaries, and feedback from accretion \citep{Tagawa:2020}.

\subsubsection{A team effort: combining multiple interactions}

Of course, the channels outlined above need not be distinct, and the same binary may have benefited from multiple types of interaction while en route to coalescence.   New studies are now attempting to consistently combine stellar and binary evolution and dynamics at comparable levels of fidelity \citep[e.g.,][]{Santoliquido:2020}.  Black holes brought together by dynamical interaction in stellar clusters may be part of mass-transferring binaries earlier in their evolutionary histories.  Mass-transferring binaries that are too wide at the end of binary evolution to merge in the age of the Universe may be helped along by subsequent dynamical encounters \citep[e.g.,][]{Belczynski:2014VMS}.  The more massive star in an isolated binary may evolve chemically homogeneously, while its less massive companion expands and donates mass \citep{Marchant:2017}.  

Hierarchical triple systems, with a tight inner binary and a third companion on a wider outer orbit, are a particularly promising combination of the isolated evolution and dynamical formation channels.  A significant fraction of massive stars appear to be born in triples and higher multiplicity systems \citep{DucheneKraus:2013, Tokovinin:2021}.  Triples may also be formed dynamically through binary--binary scatterings \citep{MillerHamilton:2002b, Trani:2021}.  Massive black holes in galactic nuclei --- or intermediate-mass black holes in globular clusters --- may also act as effective triple companions.   

When the outer binary's orbit is sufficiently wider than the inner orbit, the triple will be stable (though it may still be driven to instability by mass loss \citep[e.g.,][]{PeretsKratter:2012}).  If the inner and outer orbits in a stable triple have a high relative inclination angle, the transfer of angular momentum between these orbits can increase the eccentricity of the inner binary while keeping its semi-major axis roughly constant. For suitable initial conditions, these Lidov-Kozai oscillations in eccentricity and orbital inclination \citep{Lidov:1962,Kozai:1962} may lead to a drastic growth in the inner binary's eccentricity \citep[for a review, see][]{Naoz:2016}, reducing the binary's periapsis to the point where gravitational-wave emission can harden the binary to merger.  Recent arguments that Lidov-Kozai cycles could contribute to binary black hole mergers include \citet{SilsbeeTremaine:2017} who explored isolated triples, \citet{Antonini:2016} and \citet{Martinez:2020} in the context of dynamically formed triples in globular clusters, and \citet{Hoang:2018} who considered triples with a galactic-centre massive black hole as the outer companion.

\section{Expected properties of merging systems}\label{merge}

In this section we describe the expected hallmarks of each formation scenario in the properties of observed binary black hole mergers. These properties include merger rates, black-hole masses, mass ratios, spins, orbital eccentricities, and formation environments.

Other than the LIGO and Virgo sources, there are no known direct observations of binary black holes (except for a very speculative potential observation through microlensing \citep{Dong:2007}). As the population of gravitational-wave detections grows, so does our ability to infer merger rates and typical system properties from observed data; see the following section for a discussion of the current state of these efforts. In the meantime, our understanding of the expected merger rates and properties of binary black holes rests largely on theoretical modelling. 

For the isolated binary evolution channel (section \ref{form:isol}), this modelling typically uses a population synthesis approach, i.e., forward modelling of large populations of stellar binaries distributed according to observed initial mass and separation distributions, some (small) fraction of which end up merging as binary black holes. Population synthesis techniques can be applied either to initial binaries at formation, or to specific observed systems at intermediate evolutionary stages, such as Cygnus X-1 \citep{Bulik:2008,Neijssel:2020CygX1} and Cygnus X-3 \citep{CygnusX3:2012}.  

Models of the dynamical formation channel are typically based on either (i) computationally expensive N-body cluster models with direct integration of the equations of motion for each object, or (ii) a less computationally costly Monte Carlo approach in which the evolution of the distribution of the cluster components in phase space is coupled with stochastically sampled strong 3-body and 4-body interactions modelled with few-body dynamics, or (iii) semi-analytical prescriptions that attempt to emulate these methods at a fraction of the computational cost. 

\subsection{Merger rates}

Here we outline high-level estimates of merger rates for three candidate formation channels. The value of this back-of-the-envelope merger rate analysis is not so much in matching the rate inferred from gravitational-wave observations a posteriori as in setting the stage for some of the predictions of the properties of merging binaries (see below) and the discussion of evolutionary uncertainties (see \autoref{prospect}).  The merger rate predictions and observed rates are summarised by \cite{MandelBroekgaarden:2021} (see \cite{ratesdoc} for a historical summary from a decade earlier).  Broadly, predictions for isolated binary evolution, including contributions from chemically homogeneous evolution and population III stars, are consistent with or exceed the observed rate of binary black hole mergers of $[16,130]$ Gpc$^{-3}$ yr$^{-1}$ \citep{GWTC3:pop}, while predicted dynamical formation rates (including contributions from globular, nuclear, and young stellar clusters, as well as hierarchical triples) tend toward the lower end of the observed merger rate.

\subsubsection{Binary evolution via the common-envelope phase}\label{sec:CErates}

In order to survive an evolutionary pathway such as the one depicted in \autoref{fig:isol_binary}, a binary must (i) have the right component masses to form two black holes; (ii) have the right separation to avoid a premature merger, yet be close enough to interact; (iii) avoid disruption by supernova kicks; (iv) engage in and survive a common envelope phase; and (v) end up sufficiently compact at binary black hole formation to merge within the age of the Universe. We describe below the development of a ``Drake equation'' for the probability that an isolated stellar binary will end its life as a merging binary black hole, addressing each of these five factors in turn.   In reality, of course, the probabilities for each key stage are coupled in a rather complex manner.

\begin{enumerate}
	\item[(i)] The minimal initial stellar mass for forming a black hole is likely around $20 M_\odot$, so $f_\textrm{primary} \approx 0.1\%$ of all stars drawn from the \citet{Kroupa:2002} initial mass function will form black holes. Although this can be shifted somewhat by binary interactions, we will use this approximation for the purpose of this back-of-the-envelope calculation. Meanwhile, since the mass ratio between the secondary and the primary is roughly uniformly distributed \citep{Sana:2012,MoeDiStefano:2017}, roughly half of the secondaries will fall into the mass range of interest, $f_\textrm{secondary} \approx 0.5$.  

\item[(ii)] Initial binary  separations are observed to be distributed uniformly in the logarithm, $p(a) \propto 1/a$ \citep{Opik:1924}.  The first episode of mass transfer from the donor should typically happen when the donor has evolved beyond core hydrogen burning, but not yet beyond core helium burning.  Stars expand by several orders of magnitude during this phase, which means that the range of initial separations occupies a significant logarithmic fraction of the total initial range of separations, so $f_\textrm{init sep} \approx 0.5$.  

\item[(iii)]  We assume that black holes receive low natal kicks and do not lose significant amounts of mass during stellar collapse, so supernovae do not significantly affect binary survival or binary properties, i.e. $f_\textrm{survive SN1} \approx f_\textrm{survive SN2} \approx 1$.  

\item[(iv)] The fraction of binaries that initiate and survive the common envelope phase during mass transfer from the secondary to the primary is perhaps the least certain.  Typically, dynamically unstable mass transfer is more likely for greater mass ratios of the donor to the accretor.  The accreting black hole in this case may only have a third of the initial mass of the primary, with the rest lost with the envelope and through winds, while the secondary donor may have increased its mass during the first mass transfer phase.  Even so, unless the secondary was initially close to the primary in mass, the mass ratio is unlikely to exceed $3:1$, which may be close to the threshold for initiating a common envelope phase \citep[e.g.,][]{Claeys:2014}.  The evolutionary state of the donor during the onset of mass transfer also plays an important role, setting both the amount by which the donor shrinks in response to mass loss and the envelope binding energy.  Together, these constraints reduce the fraction of successful common envelope initiations and ejections to $f_\textrm{CE} \approx 0.1$, though this fraction depends on a range of factors including metallicity, which impacts the envelope binding energy.   

\item[(v)]  Finally, there is the question of final separation at second black hole formation, as only systems with separations smaller than those in \autoref{fig:periapsis} will merge in the age of the Universe.  The final separation after the common envelope ejection is set by the binding energy of the envelope at the time when the secondary initiates unstable mass transfer, which in turn depends on the binary separation at that time.  Therefore, very crudely, the flat-in-the-log distribution of initial binary separations persists to the final binary separation.  Since the delay time between formation and merger $\tau_\textrm{GW} \propto a^4$ is a power law in the separation, the delay time distribution also follows a flat-in-the-log distribution $p(\tau) \propto 1/\tau$.  Among the binaries that survive a common envelope the closest will be those whose separations just barely encompass a Wolf-Rayet star, i.e., with separations of a few solar radii (and merger times of a few million years).  The logarithmic distribution ensures that tens of percent of binaries that survive a common envelope will merge in the age of the Universe, $f_\textrm{merge} \approx 0.2$; this fraction is again metallicity-dependent because binaries widen more through greater wind mass loss at larger metallicities \citep{Neijssel:2019}. \end{enumerate}

Thus, our Drake-like equation is as follows:
\begin{eqnarray}
f_\textrm{BBH} &=& f_\textrm{primary} \times f_\textrm{secondary} \times f_\textrm{init sep} \times f_\textrm{survive SN1} \times f_\textrm{CE} \times f_\textrm{survive SN2} \times f_\textrm{merge} \nonumber \\
 & \sim & 0.001 \times 0.5 \times 0.5 \times 1 \times 0.1 \times 1 \times 0.2 = 5 \times 10^{-6}.
\end{eqnarray}
Merging binary black holes are rare outcomes indeed!  [Merging binary neutron stars are similarly rare: while stars with initial masses sufficient to form a neutron star, between roughly 8 and 20 solar masses, are more common than the heavier stars necessary to form black holes, neutron star natal kicks and mass loss during supernovae are more likely to disrupt binaries; the expected yields for binary neutron star mergers are within an order of magnitude of those for binary black holes.]

The star formation rate in the local Universe is $\sim 0.01 M_\odot$ Mpc$^{-3}$ yr$^{-1}$ \citep{MadauDickinson:2014}; for an average binary mass of $\sim 1 M_\odot$, the yield $f_\textrm{BBH} \sim 5 \times 10^{-6}$ corresponds to a binary black hole merger rate of $\sim 50$ Gpc$^{-3}$ yr$^{-1}$, or 5 per Myr for a Milky-Way equivalent galaxy with a space density of $0.01$ Mpc$^{-3}$.  Of course, the actual calculation of the binary merger rate requires significantly more care than the back-of-the-envelope model presented here, accounting both for initial conditions \citep{deMinkBelczynski:2015,MoeDiStefano:2017}, including orbital eccentricity, and for orbital evolution, particularly mass transfer.   This is typically undertaken by population synthesis models as described in section \ref{form:isol}.  Moreover, the redshift-dependent metallicity-specific star formation history must be convolved with the time-delay distribution between star formation and binary merger in order to determine the local merger rate \citep{Belczynski:2010,Mapelli:2017,Klencki:2018,Chruslinska:2019,Neijssel:2019,ChruslinskaNelemans:2019,Tang:2020,Broekgaarden:2021,Santoliquido:2021}.  

\subsubsection{Chemically homogeneous evolution}

We can write a similar Drake-like equation for the chemically homogeneous evolution channel.  Following \citet{MandeldeMink:2016},  the yield fraction is
\begin{eqnarray}
f_\textrm{BBH} &=& f_\textrm{primary} \times f_\textrm{secondary} \times f_\textrm{init sep} \times f_Z \times f_\textrm{merge} \nonumber \\
 & \sim & 0.0002 \times 0.5 \times 0.1 \times 0.1 \times 1 = 10^{-6}.
\end{eqnarray}

Here, $f_\textrm{primary} \approx 2 \times 10^{-4}$, lower than the value in the previous paragraph because only the most massive stars with initial masses above $\sim 50 M_\odot$ are likely to have the necessary level of mixing to undergo chemically homogeneous evolution; and $f_\textrm{secondary} \approx 0.5$ as above.  In order to engage in efficient mixing, the components must be close enough to be tidally locked and rotating at a significant fraction of the break-up velocity, yet avoid merger; this defines a range of initial separations spanning roughly a factor of 2 out of a total range spanning $\approx 5$ orders of magnitude, so $f_\textrm{init sep} \approx 0.1$ assuming a flat-in-the-logarithm distribution of initial separations.  Finally, $f_Z \approx 0.1$ is the total historical fraction of low-metallicity star formation, required to suppress line-driven winds which widen the binary, spin down the stars, and stop mixing.  Because these binaries are constructed to be very tight at birth, the vast majority should merge after the stars collapse into black holes in situ, and $f_\textrm{merge} \approx 1$. Thus, if this channel is viable, the rate of binary black hole mergers produced through the chemically homogeneous evolution channel could reach perhaps $\sim 10$ Gpc$^{-3}$ yr$^{-1}$, but with a preference for much higher masses than classical isolated binary evolution.

\subsubsection{Dynamical formation}

We can estimate an upper limit on the merger rate of dynamically formed binary black holes by assuming that all black holes in a globular cluster undergo a single merger.  About 1 in a thousand stars are massive enough to become a black hole ($f_\textrm{primary} \approx 0.001$), so even a relatively large cluster with $2\times 10^6$ stars would produce at most 2000 black holes and 1000 mergers.  [Although repeated mergers are possible if the black hole merger product is retained in the cluster and merges again, they are very rare as gravitational-wave recoil kicks are overwhelmingly likely to eject a merger remnant from the cluster \citep{Rodriguez:2018}; in any case, even merging all black holes sequentially would increase this estimate by at most a factor of 2.]  If we imagine that the merger rate is uniform over the $\sim 10$-Gyr history of the Universe (if mergers predominantly happened early in the cluster history, the rate in the local Universe would be lower), and with a globular cluster space density of around 1 per Mpc$^3$, the {\it upper limit} on the merger rate from this channel is $\sim 1000 \times 10^{-10} \textrm{yr}^{-1} \times 1 \textrm{ Mpc}^{-3} \approx 100$ Gpc$^{-3}$ yr$^{-1}$.

Young massive star clusters could add to this rate \citep[e.g.,][]{Ziosi:2014,Kumamoto:2020,Santoliquido:2020,Banerjee:2021}, as could galactic nuclear clusters \citep[e.g.,][]{MillerLauburg:2008,AntoniniRasio:2016,Hoang:2018,Stephan:2019,Grobner:2020,Tagawa:2020}.  On the other hand, in practice, the overall merger rate from the dynamical formation channel is likely to be at least a factor of a few lower for the following reasons:
\begin{enumerate}
	\item [1.]
Some single black holes will be ejected from the cluster during the three-body interactions that harden the binaries.  If the masses of all interacting objects are similar, the ejected object receives a kick that is comparable to the orbital speed of the binary.  The last scattering interaction must harden the binary to a sufficiently small separation for gravitational-wave emission to take over. The minimum orbital velocity of an equal-mass circular binary that can merge in the age of the Universe through gravitational-wave emission is $\sim 500$ km s$^{-1}$; this scales only weakly with the binary's mass, as $v \propto \sqrt{M/a}  \propto \sqrt{M/M^{3/4}} \propto M^{1/8}$ (see \autoref{fig:periapsis} and \citet{Peters:1964}).   This minimum velocity is larger than the typical globular cluster escape velocity, $\lesssim 50$ km s$^{-1}$, hence many interlopers, including some black holes, will be ejected.
	\item[2.]
Following such a 3-body interaction, the binary itself must recoil in the opposite direction from the ejected single star in order to conserve momentum. Although this recoil velocity is reduced by the ratio of the binary's mass to the interloper mass --- typically greater than $2:1$, as the lightest body is most likely ejected --- the binary may still be ejected from the cluster before it is sufficiently tight to merge through radiation reaction alone.
\item[3.] If most mergers happen soon after globular cluster formation, the local merger rate would be low because the majority of globular clusters are old \citep{Rodriguez:2016big}.
\end{enumerate}

\subsection{System Properties}
\subsubsection{Masses}
The masses of black holes in the binaries of interest will not change appreciably after the black holes are formed by stellar collapse or supernova explosions.  The first black hole to form is unlikely to accrete a significant amount of mass from its companion: doubling its mass would take more than 100 million years, as the mass-gain rate is $\dot{M}/M \lesssim 10^{-8}$ yr$^{-1}$ at the Eddington limit, and the massive companion will only survive as a non-degenerate potential donor star for a small fraction of that time (though some models allow super-Eddington accretion \cite{Eldridge:2017,Bavera:2020,vanSon:2020}). No further accretion is expected in the system after the second compact object is formed, except in very gas rich environments like active galactic nuclei. The mass of each black hole is thus almost equal to its mass at formation. 

The classical isolated binary evolution channel can produce binary black holes with a broad range of masses, matching most observations to date \citep[e.g.,][]{Stevenson:2017,Eldridge:2017,GiacobboMapelli:2018} as well as GW170817 and the observed Galactic double neutron stars \citep[e.g.,][]{Kruckow:2018,VignaGomez:2018}.  In fact, contrary to mistaken lore, the high masses of the first observed black holes were not entirely surprising: such systems were predicted to form in low-metallicity environments \citep{Mapelli:2009,Belczynski:2009} and to make a significant contribution to the detected population because more massive binaries emit more energy in gravitational waves \citep{Dominik:2014}; the effect of metallicity is discussed further in section \ref{environ} below. 

Dynamical formation in globular clusters can also lead to a range of masses, but may favour more massive binaries: lighter black holes would be ejected by heavier ones in three-body interactions \citep{Rodriguez:2015}. The mass distribution is sensitive to natal kicks, which may eject black holes at formation \citep{Zevin:2017}.  For example, \citet{Chatterjee:2017} argued that dynamical formation could explain most black-hole binary merger observations, with old, metal-poor clusters preferentially contributing more massive black-hole binaries and younger, metal-rich clusters contributing lighter binaries.
 
Chemically homogeneous evolution can only produce massive binaries, with total black-hole mass above $\sim 50 M_\odot$ \citep{MandeldeMink:2016,Marchant:2016}.

One challenge for all of these channels is explaining merging components above $\sim 50\, M_\odot$.  Models suggest that black holes in this mass range should not exist, as their progenitors would lose mass through pulsations or explode completely in pair-instability supernovae, driven by a sudden drop in the radiation pressure support as high-energy photons spontaneously produce electron-positron pairs  \citep[e.g.,][]{Woosley:2017,Farmer:2019}.  While early gravitational-wave observations appeared consistent with a maximum component mass and perhaps even an excess of $\approx 30$--$40 M_\odot$ black holes due to pulsational pair-instability supernovae \citep[e.g.,][]{Stevenson:2019}, this is no longer the case with systems like GW190521, whose components fall into the anticipated pair-instability mass gap.  Possible explanations \citep{GW190521:astro} include multiple generations of black hole mergers in clusters \citep[e.g.,][]{Rodriguez:2019,Anagnostou:2020}, additional growth through accretion, particularly in an active galactic nucleus gas disk \citep[e.g.,][]{GerosaFishbach:2021,Tagawa:2021}, or a reassessment of the location of the pair-instability mass gap \citep{Belczynski:2020massgap}.

Very asymmetric binaries, particularly GW190814, present another challenge for most evolutionary channels \citep{Zevin:2020,Mandel:2020}; one recently proposed scenario relies on the possibility that some massive progenitors may leave light remnants behind, particularly if their explosibility is enhanced by a change in compactness due to previous episodes of mass transfer \citep{Antoniadis:2021}.  

\subsubsection{Mass ratios}
The mass ratios of most merging black-hole binaries produced through the isolated binary evolution channel are likely to be on the equal side of $2:1$, because of the mass ratio constraints necessary to ensure stable mass transfer from the primary to the secondary and then dynamically unstable reverse mass transfer.   However, mass ratios exceeding $2:1$ are expected to be more common at low metallicity \citep{Dominik:2012,Stevenson:2017}, and the strict mass ratio constraint may in fact apply to black-hole binaries that only experience stable mass transfer episodes \citep{vanSon:2021}.

Dynamically formed binaries may also favour comparable masses because the heaviest black holes in a cluster are most likely to merge with each other following dynamical exchanges.  However, more extreme mass ratios are also possible if the globular cluster contains a particularly heavy stellar-mass black hole or a $\gtrsim 100 M_\odot$ intermediate-mass black hole that merges with lower-mass black holes \citep{Mandel:2008,Belczynski:2014VMS}. 

Chemically homogeneous evolution has a much stronger preference for equal masses \citep{MandeldeMink:2016}.  \citet{Marchant:2016} and \citet{Riley:2020} concluded that nearly equal-mass black holes would be produced from the evolution of systems that were in contact (shared mass) on the main sequence, before disengaging during subsequent phases of chemically homogeneous evolution and proceeding along the pathway sketched out in section \ref{form:CHE}. 

\subsubsection{Black-hole spin magnitudes and directions}\label{BHspins}
The spins of merging black holes may carry imprints of their evolutionary history \citep[e.g.,][]{GerosaBerti:2017}.  Isolated binaries, including those formed through chemically homogeneous evolution, are generally expected to have preferentially aligned spins after undergoing episodes of mass transfer and/or tidal coupling, although natal kicks and possible spin tilts during supernovae could lead to misalignment \citep[e.g.,][]{Kalogera:2000,Farr:2011,Tauris:2017}.  Meanwhile, the spin directions for dynamically formed binaries are likely to be distributed isotropically \citep[e.g.,][]{Rodriguez:2016spin}.  On the other hand, black holes merging through the Lidov-Kozai resonance in hierarchical triples could preferentially have spins in the orbital plane \citep{LiuLai:2018,RodriguezAntonini:2018}.

Spin magnitudes may also depend sensitively on evolutionary history.  A massive star may contain a large amount of angular momentum, far in excess of the maximum allowed for a spinning black hole. However, the vast majority of that angular momentum will be contained in the outer layers of the star and can therefore readily be lost through winds or via envelope stripping by a companion.  Therefore, black holes formed from stars that were rapidly spinning at some point in their evolution may still spin slowly unless the stars are spun up through mass transfer or tides shortly before collapse \citep{Kushnir:2016,HotokezakaPiran:2017,Zaldarriaga:2017}, or, less likely, during the supernova itself \citep{Batta:2017,Schroeder:2018}.   Consequently, if angular momentum transport within a star is at least moderately efficient, we may expect that the first-formed black hole in an isolated binary will always be slowly spinning, while the second-born black hole will be rapidly spinning only in the tightest binaries, when its helium star progenitor can be spun up through tides \citep{Bavera:2019,Belczynski:2020,MandelFragos:2020,Bavera:2021,OlejakBelczynski:2021}.

This argument for slowly spinning black holes in merging compact binaries is potentially in conflict with the claimed observations of rapid spins in black-hole high mass X-ray binaries \citep{FishbachKalogera:2021}.   However, there may be significant differences in the evolutionary channels and environments of the locally observed high-mass X-ray binaries and the binary black holes observed in gravitational waves \citep[][and see section \ref{environ}]{HotokezakaPiran:2017}.  In particular, black holes in high-mass X-ray binaries such as Cygnus X-1 could be spun up during mass transfer late in the primary's main sequence, when the core and envelope are still strongly coupled \citep{Valsecchi:2010,Qin:2018}.  Such systems may then be unlikely to produce merging black holes as they are too close to experience a common envelope and too wide to merge without one \citep{Neijssel:2020CygX1}, though some fraction could still merge through fortuitous natal kicks of the secondary, possibly explain the spin precession of GW151226 \citep{Chia:2021} (but see \cite{GW151226,MateuLucena:2021}).  Moreover, as discussed in section \ref{sec:XRB}, the spin magnitude measurements in X-ray binaries could be very sensitive to the modelling assumptions, and the systematic errors may dominate the statistical ones.

Meanwhile, there is active debate in the literature about the black-hole spin distribution of gravitational-wave events (see section \ref{sec:GWevents}).  Among spin variables that remain approximately constant during the inspiral, gravitational-wave spin measurements most accurately constrain the effective spin $\chi_\mathrm{eff}$ \citep{Racine:2008,Ajith:2011}, which enters the phasing of gravitational waves at the 1.5 post-Newtonian order \citep{PoissonWill:1995}.  Low values of $\chi_\mathrm{eff}$ could point to either low spin magnitudes or isotropically directed spins or a combination of both \citep{Farr:2017}, with isotropic spin directions being the hallmark of dynamical formation.  The diverging conclusions regarding the fraction of binaries with significant and preferentially aligned component spins \citep{GWTC2:pop,GWTC3:pop,Roulet:2021,Callister:2021,Galaudage:2021} appear to be a consequence of the chosen parametrised model, which may be mis-specified in some cases.  At face value, the absence of confidently negative measurements of $\chi_\mathrm{eff}$ and the roughly one fifth fraction of observed binaries with positive $\chi_\mathrm{eff}$ inconsistent with zero suggest that 80\% of observed binaries contain non-spinning black holes while 20\% have moderate net spin in the direction of the orbital angular momentum.  This appears consistent with isolated binary evolution with occasional tidal spin-up as the dominant channel.  The distribution of spin between binary components, the degree of spin precession, and possible correlations between spins and masses remain open issues. 

\subsubsection{Orbital eccentricities}\label{ecc}
	The orbital eccentricities of merging black-hole binaries may contain information about their previous evolution \citep{MandelOShaughnessy:2010}. Because gravitational waves are very efficient at damping out orbital eccentricity, isolated binaries are expected to merge on circular orbits. Dynamically formed systems may, however, retain detectable eccentricities at merger if their orbits are very tight at formation: the periapsis of a highly eccentric binary would need to be $\lesssim 10^4$ km in order for the eccentricity to exceed 0.1 when the gravitational-wave frequency reaches the minimum of the detector sensitivity band at 10 Hz.  Dynamical formation of very tight binaries may be possible through two-body captures in Galactic nuclei \citep[][but see \cite{Tsang:2013}]{OLeary:2008}; in some hierarchical triples evolving through the Lidov-Kozai mechanism \citep[e.g.,][]{AntoniniPerets:2012}; or through close captures during three-body interactions \citep{Samsing:2014, Rodriguez:2018}.  We discussed the exciting indications of possible measurements of non-zero eccentricity \citep{RomeroShaw:2021} in section \ref{sec:GWevents}.

\subsubsection{Formation environments}
\label{environ}
The evolutionary history of merging black holes also depends on their formation environments, particularly through the effect of chemical composition (metallicity) on stellar evolution and thereby black-hole mass.

Black-hole masses are set by the mass of the parent star at the end of stellar evolution, preceding the collapse or supernova explosion, and by the amount of ejecta during the supernova. Observational evidence and collapse models suggest that sufficiently massive stellar cores may completely collapse into black holes \citep[for a review, see][]{Mirabel:2016}. Thus, the mass of the stellar core --- the convective region in the centre of the star where nuclear fusion proceeds from hydrogen through helium, carbon, oxygen, and so on to iron --- determines the ultimate black-hole mass.  If left unperturbed, a third to a half of the mass of a massive star will end up in a carbon--oxygen core, and thereby in the black-hole remnant.

\begin{figure}
	\centering
	\includegraphics[width=0.8\textwidth]{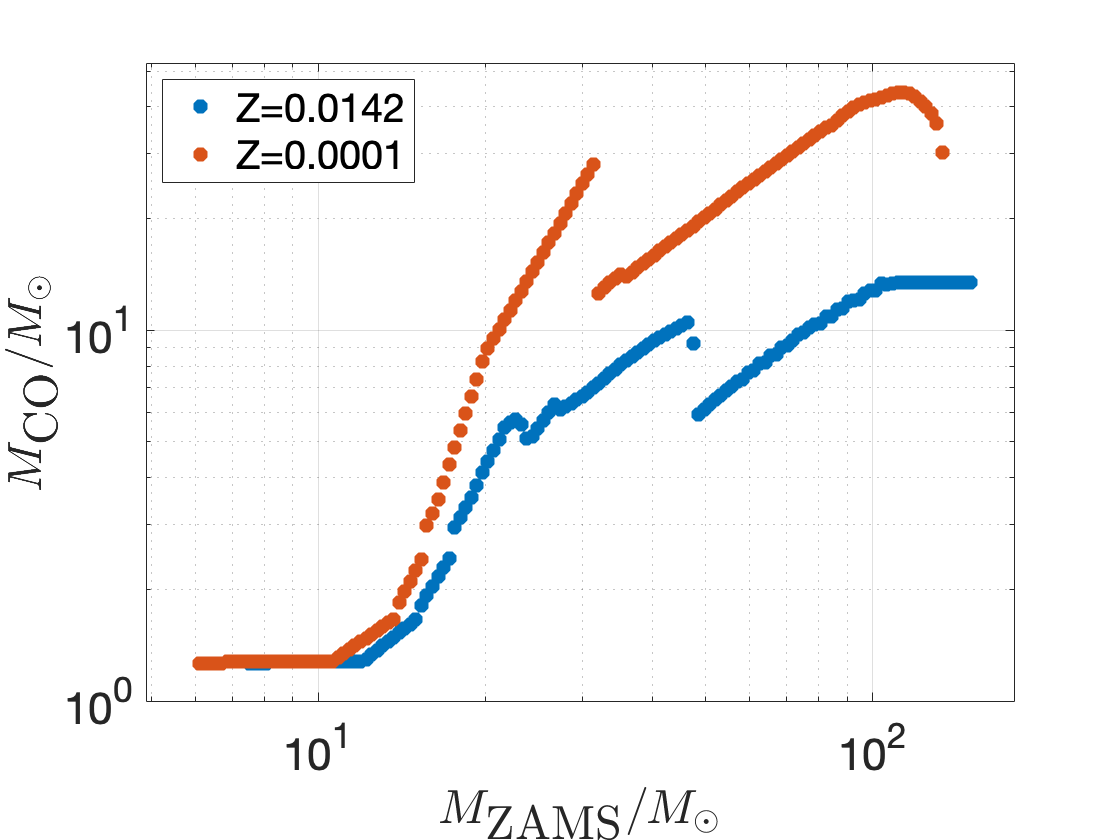}
	\caption{\label{fig:BHremnant} Remnant compact object mass from single stellar evolution for a given zero-age main sequence stellar mass, following the default COMPAS prescription for winds \citep{COMPAS:2021}  and the `delayed' \citet{Fryer:2012} supernova fallback prescription, at solar  ($Z=0.0142$, blue) and low ($Z=0.0001$, red) metallicities.  
	} 
\end{figure}

The intense radiation from massive stars with luminosities $\gtrsim 10^5 L_\odot$ drives significant winds, which can remove a large amount of material from the star. The mass of the star at the end of stellar evolution can thus be much lower than the star's initial mass, and the mass of the resulting compact object can therefore be much lower than $1/3$ of the initial stellar mass. Metals like iron, with their many absorption lines, are particularly efficient at capturing the stellar radiation and transforming it into outward momentum.  Therefore, metallicity is a key ingredient in determining stellar wind mass loss rates \citep{Vink:2001,Sander:2020}. Although wind models are uncertain \citep[e.g.,][]{Renzo:2017,Vink:2017}, simulations suggest that at the metallicity of our solar neighbourhood, massive stars lose enough mass via winds that maximum black-hole masses are only around 15--20 $M_\odot$ \citep{Belczynski:2009,Spera:2015}. This matches the masses of black holes observed in X-ray binaries. Figure \ref{fig:BHremnant} illustrates the possible dependence of compact remnant mass on metallicity. 
 
Forming heavier black holes, such as some of those observed through gravitational waves, requires stellar progenitors born in regions with lower metallicity \citep{Belczynski:2016}. The yield of merging black holes per unit star-forming mass is also greatest at lowest metallicity, due largely to the reduced wind-driven mass loss \citep[e.g.,][]{Belczynski:2010,Kruckow:2018}. For the isolated binary channel, delayed stellar expansion at lower metallicity, which allows the stars to develop a larger helium core before engaging in mass transfer, further enhances this effect \citep[e.g.,][]{Stevenson:2017}.  Most low-metallicity star formation occurred in the early Universe, before the interstellar medium was polluted with metal-rich products of stellar evolution; however, a broad range of stellar metallicities is observed at all epochs  \citep[e.g.,][]{LangerNorman:2006,TaylorKobayashi:2015}.  Of course, the mergers can be significantly delayed relative to star formation and black holes formed in the early Universe may only now be merging.  Various formation channels predict a range of delay time distributions, which could potentially provide an additional way to discriminate between them \citep{FishbachKalogera:2021time}.

\section{Prospects for gravitational-wave astronomy}\label{prospect}

The gravitational-wave events observed to date already provide a number of exciting insights into the astrophysics of binary black holes and their progenitors, but many open questions remain.  Fortunately, these observations indicate a binary black hole merger rate of around $20$ Gpc$^{-3}$ yr$^{-1}$ at redshift zero with a likely higher merger rate at higher redshifts \citep{GWTC3:pop}, which points to the prospect of hundreds of detections within the next few years, once the LIGO and Virgo instruments reach design sensitivity \citep{scenarios}.  This will yield enough information about both individually exciting events and population distributions to inaugurate gravitational-wave astronomy as a genuine tool for exploring stellar and binary evolution.  In this section, we describe some of the prospects for extracting information from the growing data set of observations over the next few years, including single informative observations, population statistical inference, and theoretical modelling.

\subsection{Individual exciting events}
Three years ago, in an earlier version of this review, we suggested that two of the most exciting individual observations could include components in either the predicted mass gap due to pair-instability supernovae or the possible mass gap between neutron stars and $\gtrsim 5\, M_\odot$ black holes.  With GW190521 and GW190814, both have now been observed!

Other individually exciting events could include the first confirmed discovery of an intermediate-mass black hole in the few-hundred solar-mass range, either as a merger of two such black holes \citep[e.g.,][]{AmaroSeoaneSantamaria:2009,Veitch:2015,Graff:2015} or as an intermediate-mass-ratio inspiral of a stellar-mass compact object into such a black hole \citep[e.g.,][]{Mandel:2008,Haster:2015IMRI,Haster:2016}. 

A source with a detectable eccentricity when gravitational-wave frequencies exceed $\gtrsim 10$ Hz, where current ground-based detectors are sensitive, would clearly signal a dynamical capture \citep[e.g.,][]{Zevin:2017,Rodriguez:2018}. 

A measurement of an effective spin value close to either 1 or $-1$ in a binary with nearly equal-mass components would point to rapid spins of both black holes. A substantial positive effective spin coupled with high masses would point to the likelihood of chemically homogeneous formation \citep{Marchant:2016}, while a substantially negative effective spin measurement would indicate an unexpected anti-alignment between rapid spins and the binary's orbital angular momentum, perhaps through either a supernova spin tilt or dynamical formation.  

Gravitational waves can be gravitationally lensed; lensing would amplify the signal and provide additional information on the gravitational-wave source \citep[e.g.,][]{Broadhurst:2019,Hannuksela:2019,Marchant:2020}.

Although electromagnetic transients are not broadly expected to be associated with binary black hole mergers \citep[e.g.,][]{Lyutikov:2016}, any such observations would point to the persistence of material around the merging binary \citep[e.g.,][]{deMinkKing:2017}.

\subsection{Population statistics}

Population statistics will generally yield more information than individual events.  We can use the statistical analysis of sufficiently large observational data sets to infer bulk properties of the source population, such as the mass and spin distributions, and to search for distinct subpopulations.  We can separate the approaches to inference on the observed populations into two types: unmodelled or weakly modelled inference, and inference that relies on specific accurate models. 

At the most basic end of weakly modelled inference is the (possibly non-parametric) reconstruction of an underlying distribution, such as the mass function of merging black holes \citep{BBH:O1,GWTC2:pop,GWTC3:pop}, while accounting for measurement uncertainties and selection effects.  For example, \citet{Fishbach:2017mass} and \citet{Edelman:2021} explored weakly parametrised mass models to search for the presence of a mass gap due to pair-instability supernovae or an excess peak due to pulsational pair-instability supernovae, while \citet{RinaldiDelPozzo:2021} considered a completely non-parametric Dirichlet process Gaussian mixture model for the mass distribution.  Meanwhile, as discussed above, several groups investigated inference on the mass and spin distributions with parametrised phenomenological models \citep[e.g.,][]{TalbotThrane:2017,GWTC2:pop,Callister:2021,Galaudage:2021,Li:2022}.  

We can look for distinct subpopulations or ``clusters'' of events in the observable parameter space, in the hope of finding distinct categories corresponding to different evolutionary channels.  These approaches have been proposed for both mass \citep{Mandel:2015,Mandel:2016cluster} and spin \citep{Farr:2018,Powell:2019} distributions.  It is worth noting that such clustering or classification schemes cannot hope to correctly assign individual events to a specific cluster, which is generally impossible given the significant measurement uncertainties \citep{Littenberg:2015}; instead, the goal is to identify the various clusters and measure the relative frequencies of events in these categories.

The next level of population-based inference relies on assuming that precise, possibly parametrised, subpopulation distributions are known.  In this case, hierarchical modelling (extreme deconvolution in the language of \cite{Hogg:2010}) can be used to simultaneously determine the ratios of different subpopulations (e.g., arising from different formation channels) and any free parameters in the subpopulation models.  Again, such approaches have been applied to both mass \citep[e.g.,][]{Zevin:2017} and spin \citep[e.g.,][]{Vitale:2015,Stevenson:2017spin} distributions and are generally expected to have greater resolving power than clustering in the absence of accurate population models.  On the other hand, reliance on precise models in the face of significant modelling uncertainties \citep[e.g.,][]{Belczynski:2021} increases the  the risk of misleading results through model misspecification.

\subsection{The future of population synthesis techniques}
There is, of course, a multitude of uncertain physics within any given model, both limiting the usefulness of approaches that rely on a precise knowledge of subpopulation distributions and, more importantly, providing a set of key science questions that we would like to answer with the aid of gravitational-wave observations.  How much mass do stars lose in winds, and what exactly is the impact of metallicity and rotation on stellar evolution \citep{LimongiChieffi:2018}?  What happens to the star's angular momentum during collapse, how much mass is ejected, and how much of an asymmetric kick does the remnant receive? How conservative is mass transfer in binaries and how much orbital angular momentum is carried away by the mass lost from the binary during non-conservative mass transfer?  What are the conditions for the onset of a common-envelope phase and common-envelope ejection, and how does the binary change in the process?   How do dynamical interactions affect binary evolution?  And, in turn, how do massive stellar binaries and their compact remnants feed back into astrophysics and cosmology on all scales?

One approach to addressing these big questions is to use the framework of population synthesis, which makes it possible to parametrise the uncertain physics and predict the expected source rates and distributions under different models.  Historically, most efforts relied on a discrete set of a few models in the parameter space of population synthesis assumptions \citep[e.g.,][]{Dominik:2012,Stevenson:2015}.  Even with the low computational cost of population synthesis, it is not feasible to explore more than  tens or hundreds of models, and this is not sufficient to cover the full parameter space or to investigate the correlations between model parameters.  However, recent successes in building accurate and computationally efficient emulators over the model parameter space \citep{Barrett:2017,TaylorGerosa:2018,Wong:2020,Lin:2021} suggest that a full exploration of this space will soon be possible. For example, \citet{Barrett:2017FIM} applied Fisher information matrix techniques to the space of model parameters and found that parameters such as those describing mass loss rates during the Wolf-Rayet and luminous blue variable phases of evolution, and common-envelope energetics, could be measured to the level of a few percent with a thousand detections. 

\subsection{Other datasets and future missions}
Additional observational datasets will further aid in interpreting gravitational-wave observations and in refining models of stellar and binary evolution.

Future gravitational-wave missions raise the prospect of observing the evolution of populations of merging binaries -- or individual systems -- across a broad band of frequencies, from the millihertz \citep[e.g.,][]{Sesana:2016} through the decihertz \citep{Mandel:2017} and hertz \citep{ET:2012}, to the LIGO/Virgo band.  The ability to track individual sources and source populations across the frequency spectrum will make it possible to combine information which can best be measured at low frequencies (e.g., eccentricity and sky location) and high frequencies (e.g., spins).  Meanwhile, a stochastic background of gravitational waves from individually unresolvable binary black holes could be measured in the next few years \citep{GW150914:stoch}, possibly providing additional insight on high-redshift populations; however, the only constraining parameter in a detection of such a background will likely be its amplitude, which carries only limited information \citep{Callister:2016}.

Ultimately, the most useful constraints are likely to come from the requirement that any candidate evolutionary model must self-consistently explain all of the available data: gravitational-wave observations, perhaps in multiple frequency bands, as well as electromagnetic observations of X-ray binaries, Galactic neutron stars, gamma ray bursts, supernovae, luminous red novae, etc. Incorporating these constraints will make it possible to resolve modelling degeneracies and to build a concordance model of massive binary evolution.  

\subsection{Outlook}

Today we have around 80 observations of merging binary black holes.  That is 80 more than 6 years ago, and we have already learned a great deal.  Binary black holes exist. They merge. They do so relatively frequently.  They have a broad range of masses. None of the sources seen so far seem to have rapid net spins in the direction opposite to the orbital angular momentum.  We also have several plausible evolutionary channels, some of which appear to explain many of the observed properties: merger rates, masses, and possibly spins. 

At the same time, we have the tools in place to create detailed models of stellar and binary evolution and of dynamical interactions under a variety of parametrisable assumptions.  Statistical techniques are in place for analysing the data and inferring properties directly from the observations or by comparison with the detailed models.  Most importantly, we have wonderful detectors which are continuing to improve, and we expect to obtain enough data in the next few years to carry out these analyses.  Perhaps in a few years we will know that all of the channels described here contribute non-negligibly to binary black hole formation; that black holes receive small kicks at birth; and that they spin slowly in merging binaries except when stripped helium stars are spun up through tides. We may also have made the first detections of new objects such as $200 M_\odot$ intermediate-mass black holes.

The future of gravitational-wave astronomy holds the thrilling prospect of addressing the inverse problem of massive binary stellar evolution: inferring the formation channels and their physics from observations of the merging compact-object binary population.  Like a palaeontologist who uses her knowledge of anatomy to determine the appearance, eating habits, and even behaviour of extinct dinosaurs from their fossilised remnants, we can now use merging black holes --- remnants of massive stars --- to probe the behaviour of those stars, and particularly their evolution in binaries.

\section*{Acknowledgements}
We would like to thank Christopher Berry, Cole Miller, Fred Rasio, Dorottya Sz\'{e}csi, Thomas Tauris, and Vicky Kalogera for comments that significantly improved this manuscript, and Lev Yungelson and Ed van den Heuvel for providing a historical perspective.  

IM is grateful to all members of Team COMPAS.  Special thanks go to Will Farr, Stephen Justham, Vicky Kalogera, Gijs Nelemans, Philipp Podsiadlowski, and especially Selma de Mink for stimulating discussions spanning many years.  Parts of this research were supported by the Australian Research Council Centre of Excellence for Gravitational Wave Discovery (OzGrav), through project number CE170100004.  IM is a recipient of the Australian Research Council Future Fellowship FT190100574.
 
AF is grateful for the astro-tourism.

Simulations in this paper made use of the COMPAS rapid binary population synthesis code (version 02.26.00), which is freely available at \url{http://github.com/TeamCOMPAS/COMPAS}.

%\section*{References}
%\bibliographystyle{hapj}
%\bibliography{Mandel}

%\bibliographystyle{cas-model2-names}
%\bibliographystyle{elsarticle-num}

\bibliography{Mandel}

\end{document}